%% file: main.tex
\documentclass[fleqn,10pt]{wlscirep}
\usepackage[utf8]{inputenc}
\usepackage[T1]{fontenc}

\definecolor{citecolor}{HTML}{0071bc}
\definecolor{citered}{HTML}{8b0000}
\usepackage{hyperref}
\usepackage{url}
\usepackage{svg}
\usepackage{bbding}

\usepackage{booktabs}
\usepackage[numbers]{natbib}
\input{math_commands.tex}

\usepackage{graphicx}
\usepackage{float}
\usepackage{amsfonts,amsmath,amssymb,amsthm,mathtools}
\usepackage{cleveref}
\usepackage{multicol, multirow}
\usepackage{makecell}
\usepackage{caption}
\usepackage{subfigure}
\usepackage{adjustbox}
\usepackage{enumitem}
\usepackage{wrapfig}
\usepackage{algorithm}
\usepackage{algorithmic}
\usepackage{authblk}
\usepackage{afterpage}
\usepackage{dsfont}
\usepackage{diagbox}
\usepackage{arydshln}
\usepackage{threeparttable}
\usepackage{graphicx}
\usepackage{subcaption}
\usepackage{graphicx}
\usepackage{subcaption}
\usepackage{amssymb}
\usepackage{pifont}
\usepackage{diagbox}
\usepackage{multirow}
\usepackage{tikz}

\usepackage{bbm}
\usepackage{accents}
\usepackage{cleveref}
\usepackage{bbding}
\usepackage{geometry} 
\usepackage{tabularx}

\newcommand{\ie}{\em{i.e.}}
\newcommand{\eg}{\em{e.g.}}

\usepackage{xcolor}
\definecolor{ForestGreen}{RGB}{34,139,34}

\usepackage{xcolor}
\newcommand{\model}[1]{NucleusDiff}
\definecolor{c1}{HTML}{E83100}
\definecolor{c2}{HTML}{2F70AF}
\definecolor{c3}{HTML}{c40f40}

\usepackage{manyfoot}

\DeclareNewFootnote{default} 
\DeclareNewFootnote{A}[alph] 
\DeclareNewFootnote{B}[Alph] 
\DeclareNewFootnote{C}[roman] 
\DeclareNewFootnote{D}[Roman] 
\DeclareNewFootnote{E}[fnsymbol]

\title{Manifold-Constrained Nucleus-Level Denoising Diffusion Model for Structure-Based Drug Design}
\author[1,2]{Shengchao Liu$^*$}
\author[1]{Divin Yan$^*$}
\author[3]{Weitao Du}
\author[4]{Weiyang Liu}
\author[5]{Zhuoxinran Li}
\author[6]{Hongyu Guo}
\author[2]{Christian Borgs$^\dagger$}
\author[2]{Jennifer Chayes$^\dagger$}
\author[1]{Anima Anandkumar$^\dagger$}
\affil[1]{California Institute of Technology}
\affil[2]{University of California Berkeley}
\affil[3]{Alibaba DAMO Academy}
\affil[4]{Max Planck Institute for Intelligent Systems, T\"ubingen}
\affil[5]{University of Toronto}
\affil[6]{University of Ottawa}

\begin{abstract}
Artificial intelligence models have shown great potential in structure-based drug design, generating ligands with high binding affinities. However, existing models have often overlooked a crucial physical constraint: atoms must maintain a minimum pairwise distance to avoid separation violation, a phenomenon governed by the balance of attractive and repulsive forces. To mitigate such separation violations, we propose NucleusDiff. It models the interactions between atomic nuclei and their surrounding electron clouds by enforcing the distance constraint between the nuclei and manifolds. We quantitatively evaluate NucleusDiff using the CrossDocked2020 dataset and a COVID-19 therapeutic target, demonstrating that NucleusDiff reduces violation rate by up to 100.00\% and enhances binding affinity by up to 22.16\%, surpassing state-of-the-art models for structure-based drug design. We also provide qualitative analysis through manifold sampling, visually confirming the effectiveness of NucleusDiff in reducing separation violations and improving binding affinities.
\end{abstract}

\begin{document}
\flushbottom
\maketitle
\thispagestyle{empty}
\def\thefootnote{$*$}\footnotetext{Equal contribution.}
\def\thefootnote{$\dagger$}\footnotetext{Joint supervision.}

\section{Introduction}
Structure-based drug design is a cornerstone of drug discovery, aiming at designing small molecule ligands based on the geometric structures of biological targets, typically the protein pockets. It faces significant challenges due to the vast chemical search space and the complex geometric interactions between ligands and proteins in three-dimensional Euclidean space~\cite{johnson2013druggable}. To cope with these challenges, machine learning has emerged as a powerful tool for efficiently navigating the design space of small molecules and effectively generating ligands with high binding affinities.

\begin{figure*}[t]
\centering
\includegraphics[width=\linewidth]{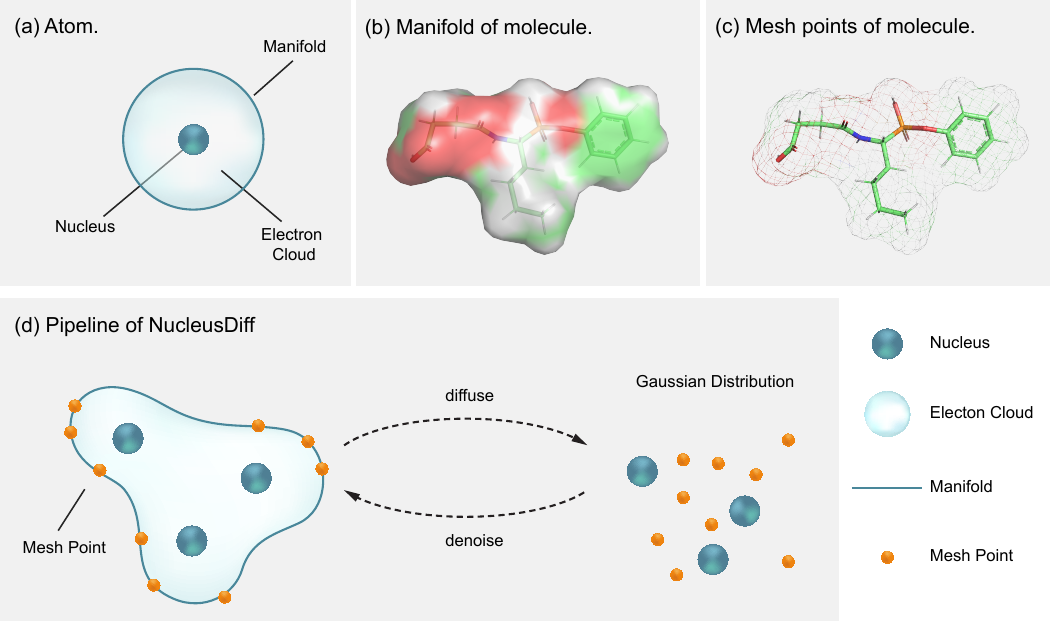}
\vspace{-4ex}
\caption{\small
(a) Illustration of the nucleus, the electron cloud, and the manifold of an atom. The electron cloud represents the probabilistic distribution of electrons around the nucleus, and the manifold is the sphere corresponding to the average distance from the nucleus to the outermost electrons in the electron cloud. (b) Illustration of the manifold surrounding a molecule. (c) Illustration of the mesh points obtained from discretizing a manifold. (d) Pipeline of \model{}. \model{} performs denoising diffusion on both the nuclei and the discretized mesh points, where the distances between them approximate the van der Waals radii.
}
\label{fig:pipeline}
\vspace{-2ex}
\end{figure*}

\begin{figure*}[ht]
\centering
\includegraphics[width=\textwidth]{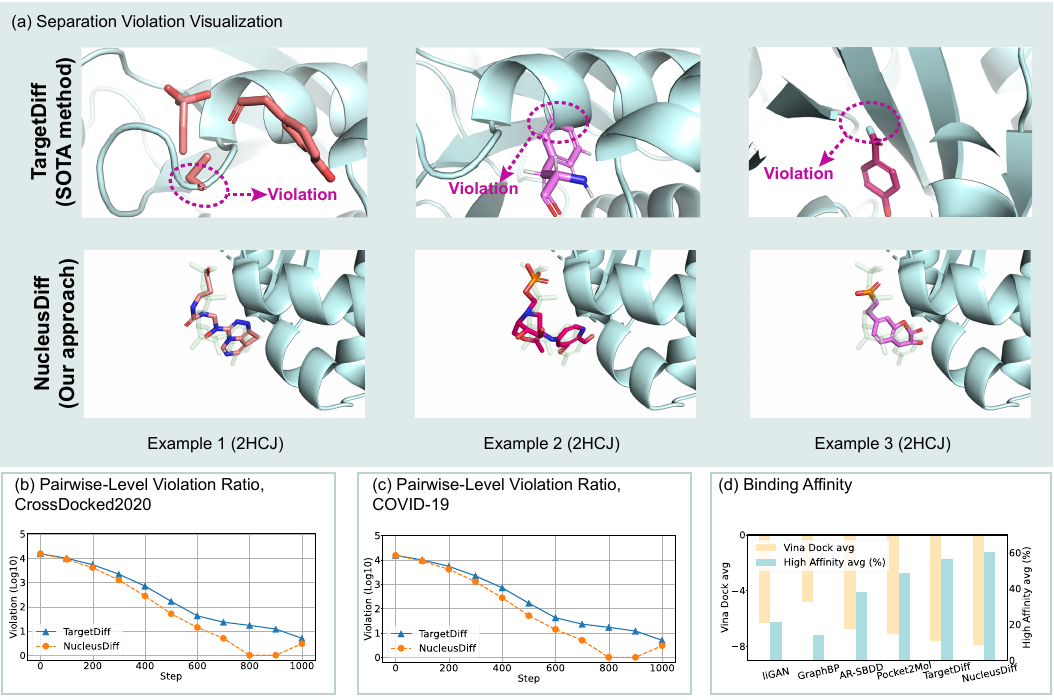}
\vspace{-4ex}
\caption{\small
(a) Visualization of generated ligands for the target 2HCJ. (b) Visualization of the pairwise-level violation ratio in TargetDiff and \model{} during the inference on the CrossDocked2020 dataset. (c) Visualization of the pairwise-level violation ratio in TargetDiff and \model{} during the inference on the COVID-19 therapeutic target. (d) Visualization of the binding affinities (in Vina Dock) for 10k sampled ligands and given proteins.
}
\label{fig:atomic_collision_illustration}
\end{figure*}

State-of-the-art deep generative models for the task generate ligands with high binding affinities by leveraging physical properties, such as the equivariance of molecular systems to rotations and translations~\cite{thomas2018tensor,liu2023symmetryinformed,luo20213d,liu2022graphbp,peng2022pocket2mol,guan20223d,abramson2024accurate}. However, they suffer from certain limitations due to the approximations in modeling. For instance, they treat each atom as a solid point, which is inconsistent with the reality that each atom occupies a distinct sphere around the atomic nucleus, known as the electron cloud. The electron cloud is a probability distribution describing the likely locations of electrons, with their behaviors governed by attractive forces ({\eg} between electrons and protons in the nuclei) and repulsive forces ({\eg}, electron pair repulsion). As illustrated in~\Cref{fig:pipeline}, these factors collectively influence the distribution and arrangement of electrons, adhering to a physical law that requires atoms to maintain a minimum pairwise distance to avoid separation violation. Having overlooked this principle, current deep generative models could breach fundamental physical laws and result in separation violation, where two atoms in the generated ligand-pocket pairs are positioned too close to each other, as depicted in~\Cref{fig:atomic_collision_illustration}.

One straightforward approach to address this issue is to consider separation violations into deep learning models as a regularization term. However, since this is an atom pairwise measure, the computational complexity scales quadratically with the number of atoms. Therefore, a more efficient and effective method is necessary to address the separation violation problem.

\textbf{Our Contributions.}
To tackle this challenge, we propose \model{}, a manifold-constrained denoising diffusion model for structure-based drug design. To incorporate inductive biases from quantum physics, \model{} directly models the constraints between atomic nuclei and their surrounding electron clouds to prevent separation violation. In essence, \model{} employs two sub-modules: one for modeling the atomic nuclei and the other one for modeling the manifold, a sphere corresponding to the average distance (a.k.a. van der Waals radius) from the nuclei to the outermost electrons in the electron cloud. Then \model{} leverages a regularization term to align the distance between nuclei and sampled manifold points with the van der Waals radii. This alignment implicitly maintains proper pairwise atomic distances and upholds the principles of attraction and repulsion that govern molecular interaction. Additionally, the complexity of modeling such a manifold constraint increases linearly to the number of atoms, making \model{} an efficient paradigm. 

We verify the effectiveness of \model{} using 100K protein-ligand binding complexes from the CrossDocked2020~\cite{francoeur2020three} firstly. Our quantitative analysis demonstrates that \model{} significantly outperforms the state-of-the-art models. \model{} demonstrates promising performance by achieving a 22.16\% improvement in binding affinity (Vina Score) compared to the previous state-of-the-art method (TargetDiff~\cite{guan20223d}) on CrossDocked2020. Owing to our design,  \model{} effectively minimizes separation violations, achieving an almost negligible violation rate, with reductions reaching up to 100.00\%. Additionally, our case study on the COVID-19 therapeutic target shows that \model{} not only achieves a superior binding affinity, with an improvement of up to 21.37\%, but also reduces the violation rate by up to 66.67\%.

In summary, the proposed \model{} reaches an optimal balance in maintaining proper physical distances to avoid separation violation and preserving the critical biochemical properties including binding affinity and stability. We believe this work will shed light on integrating physical laws into generative models for structure-based drug discovery, establishing a new direction in the biochemistry and machine learning communities.\looseness=-1

\section{Results}

\subsection{Background}
\paragraph{Small Molecules and Proteins.}
In this work, we consider small molecule ligands, which are sets of atoms in the 3D Euclidean space, $\{\vv^L, \vx^L\}$, where $\vv^L$ represents the atomic number and $\vx^L$ represents the atomic nucleus coordinate. Proteins are macromolecules, {\ie}, chains of residues. In nature, there are 20 types of residues. Each residue is a small molecule, with a fixed backbone structure: a basic amino group, an acidic carboxyl group, a side chain that is unique to each amino acid, and a carbon $C_\alpha$ connecting three components. In this work, we consider modeling proteins in the backbone-level information, {\ie}, the backbone atomic number and backbone atomic nucleus coordinates, $\{\vv^P, \vx^P\}$.\looseness=-1

\paragraph{Nucleus and Electron Cloud.}
Each atom constitutes a nucleus and an electron cloud surrounds each nucleus, as shown in~\Cref{fig:pipeline}. Recent works employ manifold learning on such an electron cloud for molecule property prediction~\cite{wang2022learning, zhang2023learning} and protein modeling in structure-based drug design~\cite{mallet2023atomsurf}. In this work, we consider modeling the manifold around each nucleus in the ligands, and the approximated distance is the van der Waals radii. Then we discretize the manifold into triangle mesh points, a form suitable for computational analysis. This is implemented using the Python package PyMesh~\cite{zhou2019pymesh}. For notation, for each ligand atom $(\vv^L, \vx^L)$, the coordinate and type of nuclei are the same as the atom-level, {\ie}, $\vv^N \triangleq \vv^L$ and $\vx^N \triangleq \vx^L$. The coordinate of the discretized points on the manifold is marked as $\vx^M$. Notice that we use a special token to delegate the electron points on the manifold.

\paragraph{Structure-based Drug Design.}
The structure-based drug design (SBDD) task utilizes the geometric structures of proteins to design and optimize ligands, like small molecules. This can be formulated as a conditional distribution modeling problem, $p(\vx^L, \vv^L |\vv^P, \vx^P)$. Notice that \model{} improves this objective function by introducing nucleus-level modeling combined with manifold-sensitive constraints of small molecules, and the problem formulation becomes $p(\vx^N, \vv^N, \vx^M |\vv^P, \vx^P)$. More details will be discussed in the Materials and Methods Section.\looseness=-1

\subsection{Separation Violation and Violation Ratio Metrics} \label{sec:collision measurement}
The separation violation occurs when two atoms come too close to each other, contradicting the underlying physical laws. We introduce the covalent radius $d$ to measure it, as the covalent radius is a more strict quantity involving the covalent bonding. Suppose we have one ligand atom coordinate $\vx_i$, one protein atom coordinate $\vx_j$, and the corresponding covalent radii are $d_i$ and $d_j$, respectively. Then during the sampling stage for ligand generation, if two atoms get too close to each other, {\ie}, $\| \vx_j - \vx_j \| \le D_{ij} = d_i + d_j$, we refer to this as the separation violation problem. To quantify this in existing deep generative models, we propose three violation ratio metrics from three granularities. For clarity, we only present one of the representative metrics in the main article.

\paragraph{Pairwise-Level Violation Ratio~(PLVR).}
The metric is the atom pairwise-level violation ratio (PLVR), which quantifies the violation ratio between all the ligand atoms and protein atoms. For each ligand atom ($\vx^L_i$), we extract its $K$ nearest protein atoms within the binding site. Then the PLVR is defined as
\small{
\begin{equation}
\begin{aligned}
    \text{PLVR} = \frac{\sum_{k \in N_{\text{mol}}, i \in N^k_{\text{atom}}, j \in N^i_{\text{nearest}}} \mathbbm{1} (\| \vx_i - \vx_j \| < D_{ij})}{K \cdot \sum_{k \in N_{\text{mol}}}N^k_{\text{atom}}},
\end{aligned}
\end{equation}
}
where $N_{\text{mol}}$ is the number of ligand molecules, $N^k_{\text{atom}}$ is the number of atoms in the $k$-th molecule, $N_{\text{nearest}}^{i}$ is the number of the nearest protein atoms of the $t$-th ligand atom, and $\mathbbm{1}(\cdot)$ is the indicator function.

We present the other two metrics in the supplementary materials. The proposed violation ratio metrics assess the separation violations between pockets and generated ligands, enhancing our understanding of the generative model's inference process for structure-based drug design. We initially test them on the generated ligands by a state-of-the-art generative model~\cite{guan20223d}. As illustrated in~\Cref{fig:atomic_collision_illustration}, existing works exhibit the separation violation issue to a certain extent.

To overcome this violation, these metrics can be directly incorporated into the training objective when modeling, but the computational complexity of these metrics is as high as $O(N_{mol} N_{atom}^2)$, where $N_{mol}$ is the number of molecules and $N_{atom}$ is the combined number of atoms in the molecule and the protein pocket. Subsequently, we introduce \model{}, an efficient model with higher binding affinities and lower separation violations.

\subsection{Manifold-Constrained Nucleus-Level DDPM: \model{}}
We propose \model{} to reduce the separation violation in the deep generative model for the structure-based drug design. The main idea is to jointly model the atomic nucleus and the manifold of the electron cloud using a denoising diffusion model. In this section, we provide a brief introduction to \model{}, and more detailed descriptions can be found in the Materials and Methods section.

\paragraph{Diffusion Model for Geometry Generation.}
We first introduce the denoising diffusion model for density estimation on general geometries, $\vx$. The denoising diffusion probabilistic model~(DDPM) consists of two stages: a forward and a backward process~\cite{ho2020denoising}. The forward process gradually adds noise to the input geometric data $\vx_0 = \vx$ to a prior Gaussian distribution $\vx_T$, and the backward process is the denoising process from the prior distribution to the data distribution. In concrete, suppose the data distribution is $\vx \sim q(\vx)$, and we have $T$ forward and backward steps with the scheduled variance $\{\beta_t\}_{t=1}^T$. Then each forward step can be represented as $q(\vx_{t} | \vx_{t-1}) = \mathcal{N}(\vx_t; \sqrt{1 - \beta_t} \vx_{t-1}, \beta_t I)$, which gives $q(\vx_t | \vx_0) = \mathcal{N}(\vx_t; \sqrt{\bar \alpha_t} \vx_0, (1-\bar \alpha_t) I)$, where $\alpha_t = 1 - \beta_t$ and $\bar \alpha_t = \prod_{s=1}^t \alpha_s$. Following the Bayes theorem, the posterior $q(\vx_{t-1} | \vx_t, \vx_0)$ can also be expressed as a Gaussian distribution:
\begin{equation}
q(\vx_{t-1} | \vx_t, \vx_0) = \mathcal{N}(\vx_{t-1}; \tilde \mu(\vx_t, \vx_0), \tilde \beta_t I),
\end{equation}
where 
$\tilde \mu(\vx_t, \vx_0) = \frac{\sqrt{\bar \alpha_{t-1}} \beta_t}{1 - \bar \alpha_t} \vx_0 + \frac{\sqrt{\alpha_t} (1 - \bar \alpha_{t-1}}{1-\bar \alpha_t} \vx_t$ and $\tilde \beta_t = \frac{1 - \bar \alpha_{t-1}}{1 - \bar \alpha_t}$.
The goal is to maximize the log-likelihood of data distribution $p(\vx)$, and after reparameterization, we aim to directly predict the ground-truth coordinates $\vx_0$ with a parameterized network $\hat \vx_0 = \phi_\theta(\vx_t, t)$. The training objective is
\begin{equation} \label{eq:kl_divergence_in_DDPM}
\begin{aligned}
\mathcal{L}_{t-1}(\vx) = \mathbb{E}_q \big[ \| \vx_0 - \hat \vx_0  \|^2 \big].
\end{aligned}
\end{equation}
Please refer to the DDPM paper for detailed derivations~\cite{ho2020denoising}.

\Cref{eq:kl_divergence_in_DDPM} holds for arbitrary density estimation, and in the following paragraphs, we will discuss how we adapt this for our proposed \model{} for structure-based drug design.

\paragraph{Nucleus-Level Diffusion Model for Ligand Generation.}
In our task, the goal is to model the atomic types and coordinates in ligands given the pocket structure: $p(\vv^L, \vx^L|\vv^P, \vx^P)$. The training objective includes a categorical term $\mathcal{L}_{t-1}(\vv^L)$ on atomic types and a continuous term $\mathcal{L}_{t-1}(\vx^L)$ on atomic coordinates~\cite{guan20223d}. Recall that with loss of generality, we can interchange atom information (type and coordinate) with nucleus information (type and coordinate), {\ie}, $\vv^N = \vv^L$ and $\vx^N = \vx^L$.

In this paragraph, we mainly discuss the continuous objective $\mathcal{L}_{t-1}(\vx^N)$, while the discrete objective function $\mathcal{L}_{t-1}(\vv^N)$ over atomic type is described in the Materials and Methods section. Then adopting \Cref{eq:kl_divergence_in_DDPM} for atomic coordinates, the training objective becomes
\begin{equation} \label{eq:kl_divergence_in_DDPM_nuclei}
\begin{aligned}
\mathcal{L}_{t-1}(\vx^N) &= \mathbb{E}_q \big[ \| \vx^N_0  - \hat \vx^N_0 (\vx^N_t, t, \vv^P, \vx^P) \|^2 \big].
\end{aligned}
\end{equation}
We note that the parameterized network $\hat \vx^N_0$ should be equivariant to the rotations and translations of the whole molecular system~\cite{vincent2011connection,song2019generative,ho2020denoising}. On the other hand, recent works~\cite{abramson2024accurate} claim that such SE(3)-equivariant modeling is unnecessary for achieving high binding affinities in structure-based drug design tasks. Meanwhile, the manifold constraint introduced in \model{} is agnostic to the design of the parameterized network, and we follow the state-of-the-art method in this work~\cite{guan20223d}.

\paragraph{Manifold-Constrained Nucleus-Level Diffusion Model for Ligand Generation.}
To reduce the separation violation, we introduce an extra constraint term that keeps the distance between the nucleus and the manifold over the electron cloud as van der Waals radius, $R$. This constraint aligns with the fundamental physical principles, as electron clouds exert both attraction and repulsion forces, preventing separation violation. To adopt this into modeling, for each nucleus, we obtain its $K$ closest mesh points in the manifold, marked as $\vx^M$. Thus, the goal becomes to maximize the joint distribution of nuclei and mesh points conditioned on the pocket, as $p(\vv^N, \vx^N, \vx^M | \vv^P, \vx^P)$. To adapt this into~\Cref{eq:kl_divergence_in_DDPM}, the objective of the sampled mesh points is 
\small{
\begin{equation} \label{eq:kl_divergence_in_DDPM_mesh_points}
\begin{aligned}
\mathcal{L}_{t-1}(\vx^M) & = \mathbb{E}_q \big[ \| \vx^M_0 - \hat \vx^M_0(\vx^M_t, t, \vv^P, \vx^P) \|^2 \big].
\end{aligned}
\end{equation}
}
On the other hand, recall that the mesh points are scattered around the nuclei with a fixed van der Waals radius $R$. Thus we add a regularization term by forcing the distance between each corresponding mesh point $\vx^N_j$ and nucleus $\vx^M_i$ to be close to $R$:
\small{
\begin{equation} \label{eq:regularization_term}
\begin{aligned}
&\mathcal{L}_{t-1}(\vx^N, \vx^M, R) = \sum_{i} \sum_{j} \\
&\quad\quad \| \| \hat \vx^N_0 (\vx^N_t, t, \vv^P, \vx^P) - \hat \vx^M_0(\vx^M_t, t, \vv^P, \vx^P) \| - R_{ij} \|.
\end{aligned}
\end{equation}
}
We call such a manifold-constrained nucleus-level modeling as \model{}. The overall objective function of the \model{} is:
\small{
\begin{equation} \label{eq:overall_objective}
\begin{aligned}
\mathcal{L} = & \mathbb{E}_{t} [ \mathcal{L}_{t-1}(\vx^N) + \mathcal{L}_{t-1}(\vv^N) + \\
& \quad\quad \mathcal{L}_{t-1}(\vx^M) + \mathcal{L}_{t-1}(\vx^N, \vx^M, R)].    
\end{aligned}
\end{equation}
}

\subsection{Experimental Setup} \label{sec:experiments}
\paragraph{Datasets.}
We conduct experiments on two datasets. (1) We first utilize CrossDocked2020~\cite{francoeur2020three} to train and evaluate our model. Following the approach of~\cite{luo20213d}, we refine the dataset of 22.5 million docked protein-ligand binding complexes by selecting the poses with a root-mean-square deviation (RMSD) of less than 1\,\AA\ and protein sequence identity below 30\%. Ultimately, we have 100,000 complexes for training and 100 complexes for testing. (2) We also carry out experiments on the COVID-19 target. Additionally, we assess the generalization abilities of \model{} on two other real-world therapeutic targets, as detailed in the supplementary materials.\looseness=-1

\paragraph{Mesh Point Construction.}
We utilize MSMS~\cite{ewing2010msms} to compute the solvent-excluded surface of the molecule. It generates a triangular mesh data structure for small molecules, and we set the argument probe radius to 1.5\,\AA\ and sampling density argument to 3.0. MSMS possesses certain challenges such as degenerate vertices and disconnected surfaces, which can disrupt the uniform distribution of mesh points. We employ PyMesh to overcome these issues~\cite{zhou2019pymesh}. It enhances the mesh quality by reducing the vertex count and correcting errors in poorly meshed regions. Finally, we select the $K$ mesh points that are closest to the van der Waals radius from the nucleus to construct a mesh point dataset. Notably, this mesh point data structure, which predominantly includes the 3D coordinates of the mesh points, is only required during the learning phase and not during inference.

\paragraph{Baselines.}
For benchmarking, we compare with various baselines: liGAN~\cite{ragoza2022chemsci}, AR~\cite{luo20213d}, Pocket2Mol~\cite{peng2022pocket2mol}, GraphBP~\cite{liu2022graphbp} and TargetDiff~\cite{guan20223d}. liGAN~\cite{ragoza2022chemsci} is a 3D CNN-based method that generates 3D voxelized molecular images following a conditional VAE scheme. AR~\cite{luo20213d}, Pocket2Mol~\cite{peng2022pocket2mol}, and GraphBP~\cite{liu2022graphbp} are GNN-based methods that generate 3D molecules by sequentially placing atoms into the protein binding pocket. We choose AR-SBDD~\cite{luo20213d} and Pocket2Mol~\cite{peng2022pocket2mol} as representative baselines due to their outstanding performance. TargetDiff~\cite{guan20223d} is the state-of-the-art model in this research line, and it employs a diffusion-based technique for generating atom coordinates and types.\looseness=-1

\begin{table*}[t]
\centering
\caption{\small
A summary of 14 biochemical properties for reference ligands and molecules generated by baseline models and \model{}. The symbols ($\uparrow$) and ($\downarrow$) indicate a higher or lower value of the metric is preferable, respectively. `Avg.' and `Med.' represent average and median values, respectively. Due to incompatibility between certain atom types produced by liGAN~\cite{ragoza2022chemsci} and GraphBP~\cite{liu2022graphbp} and the parsing capabilities of AutoDock Vina, we employ QVina~\cite{alhossary2015fast} to conduct the docking simulations for these two methods.}
\label{tab:mol_prop}
\vspace{-2ex}
\begin{adjustbox}{max width=\textwidth}
\begin{tabular}{c cc cc cc cc cc cc cc}
\toprule[1.5pt]
\multirow{2}{*}{\textbf{Metrics}} & \multicolumn{2}{c}{\textbf{Vina Score ($\downarrow$)}} & \multicolumn{2}{c}{\textbf{Vina Min ($\downarrow$)}} & \multicolumn{2}{c}{\textbf{Vina Dock ($\downarrow$)}} & \multicolumn{2}{c}{\textbf{High Affinity ($\uparrow$)}} & \multicolumn{2}{c}{\textbf{QED ($\uparrow$)}}   & \multicolumn{2}{c}{\textbf{SA ($\uparrow$)}} & \multicolumn{2}{c}{\textbf{Diversity ($\uparrow$)}} \\
\cmidrule(lr){2-3} \cmidrule(lr){4-5} \cmidrule(lr){6-7} \cmidrule(lr){8-9} \cmidrule(lr){10-11} \cmidrule(lr){12-13} \cmidrule(lr){14-15} 
 & \textbf{Avg.} & \textbf{Med.} & \textbf{Avg.} & \textbf{Med.} & \textbf{Avg.} & \textbf{Med.} & \textbf{Avg.} & \textbf{Med.} & \textbf{Avg.} & \textbf{Med.} & \textbf{Avg.} &\textbf{Med.} & \textbf{Avg.} & \textbf{Med.}\\
\midrule
\textbf{Reference}   & -6.36 & -6.46 & -6.71 & -6.49 & -7.45 & -7.26 & -  & - & 0.48 & 0.47 & 0.73  & 0.74  & -  & -    \\
\textbf{liGAN} $^*$       & - & - & - & - & -6.33 & -6.20 & 21.1\% & 11.1\% & 0.39 & 0.39 & 0.59 & 0.57 & 0.66 & 0.67   \\
\textbf{GraphBP} $^*$     & - & - & - & - & -4.80 & -4.70 & 14.2\% & 6.7\% & 0.43 & 0.45 & 0.49 & 0.48 & \textbf{0.79} & \textbf{0.78} \\
\textbf{AR-SBDD}          & \underline{-5.75} & -5.64 & -6.18 & -5.88 & -6.75 & -6.62 & 37.9\% & 31.0\% & \underline{0.51} & \underline{0.50} & \underline{0.63} & \underline{0.63} & 0.70 & 0.70  \\
\textbf{Pocket2Mol}  & -5.14 & -4.70 & -6.42 & -5.82 & -7.15 & -6.79 & 48.4\% & 51.0\% & \textbf{0.56} & \textbf{0.57} & \textbf{0.74}  & \textbf{0.75} & 0.69 & 0.71\\
\textbf{TargetDiff} & -5.01   &\underline{-5.69}  & \underline{-6.33}   & \underline{-6.47} &\underline{-7.62}  & \underline{-7.64} & \underline{56.3\%}  &\underline{57.3\%} & 0.48     & 0.48 &0.59  &   0.58  & 0.71    & 0.71      \\
\textbf{\model{} (ours)} &\textbf{-6.12}  &\textbf{-6.80}  &\textbf{-6.93}  &\textbf{-6.85}  & \textbf{-7.90} & \textbf{-7.76}  & \textbf{60.1\%} & \textbf{63.0\%}    & 0.39     &0.39  & 0.53 &0.53   & \underline{0.74}&\underline{0.73}  \\
\bottomrule[1.5pt]
\end{tabular}
\end{adjustbox}
\end{table*}

\begin{table}[t!]
\centering
\caption{\small
The PLVR performance among pocket-ligand pairs for structure-based drug design in the CrossDocked2020 and COVID-19 target.
}
\label{tab:collision_plvr}
\vspace{-2ex}
\begin{adjustbox}{max width=\textwidth}
\begin{tabular}{l lll lll}
\toprule[1.5pt]
\multirow{2}{*}{\textbf{Models}} & \multicolumn{2}{c}{\textbf{CrossDocked2020}} & \multicolumn{2}{c}{\textbf{COVID-19 Target}} \\
\cmidrule(lr){2-3} \cmidrule(lr){4-5}
 & \textbf{TargetDiff} & \textbf{\model{}} & \textbf{TargetDiff} & \textbf{\model{}} \\
\midrule
\textbf{Step-700}  & 78/2300930 & \textbf{7/2300930}  & 23/210000 & \textbf{5/210000} \\
\textbf{Step-800}  & 77/2300930 & \textbf{2/2300930}  & 17/210000 & \textbf{1/210000} \\
\textbf{Step-900}  & 78/2300930 & \textbf{4/2300930}  & 12/210000 & \textbf{1/210000} \\
\textbf{Step-1000} & 65/2300930 & \textbf{0/2300930}  & 05/210000  & \textbf{3/210000} \\
\bottomrule[1.5pt]
\end{tabular}
\end{adjustbox}
\end{table}

\subsection{Separation Violation Evaluation on CrossDocked2020}
\paragraph{The Violation Metrics.} 
To gain a comprehensive understanding of how the separation violation problem evolves during the inference process of diffusion-based probabilistic models~\cite{ho2020denoising} for structure-based drug design, we compare \model{} with the state-of-the-art model in this research line, TargetDiff~\cite{guan20223d}.
For both methods, we use 1000 timesteps for training and inference. We analyze the PLVR metric for separation violation at 11 timesteps, sampled every 100 intervals from Step-0 to Step-1000. We present the separation violation ratios from Step-700 to Step-1000 in~\Cref{tab:collision_plvr}, with the complete results available in the supplementary materials.

\paragraph{Analysis.}
As shown in~\Cref{tab:collision_plvr}, TargetDiff maintains stable separation violation ratios from Step-700 to Step-1000. In contrast, \model{} demonstrates consistently lower separation violation ratios during these inference steps. Notably, \model{} outperforms TargetDiff by nearly an order of magnitude in terms of the PLVR metric. In the final sampling step, \model{} achieves an almost negligible violation ratio, further demonstrating its superior performance. As illustrated in~\Cref{fig:atomic_collision_illustration}, the convergence trends of \model{} and TargetDiff differ markedly when evaluated using the PLVR metric, with \model{} demonstrating a significantly more pronounced convergence. More comprehensive results and analyses of \model{}'s separation violation performance are provided in the supplementary materials.

\subsection{Binding Affinity Evaluation on CrossDocked2020} 
\paragraph{The General Metrics.}
The Vina Score, Vina Min, and Vina Dock metrics are employed to evaluate the binding affinity and potential biological efficacy of small molecule drug candidates in interaction with target proteins, via the computation of docking efficiency scores. Following the methodologies outlined in \cite{luo20213d, peng2022pocket2mol, guan20223d}, we utilize the open-source AutoDockTools software~\cite{morris2009autodock4} for these calculations. The High Affinity metric gauges the strength of the ligand-target protein interaction.

Additionally, we follow existing works and consider three more metrics~\cite{luo20213d, peng2022pocket2mol, guan20223d}. QED provides a numerical assessment of a compound's drug-like characteristics, with higher values indicating a greater propensity for a compound to embody successful drug attributes. SA quantifies the ease with which a compound can be synthesized. Lastly, Diversity measures the range and heterogeneity of molecular structures and properties across a set of compounds.

\paragraph{Analysis.}
We generate 100 ligand molecules for each protein target in the test set, resulting in a total of 10,000 molecules. The size of each generated molecule, {\ie}, the number of atoms in each molecule, is determined by sampling from the size distribution observed in the training set. The comprehensive results for \model{} and the baselines are displayed in~\Cref{tab:mol_prop}.

We note that \model{} surpasses all baseline models on 8 out of 14 evaluated metrics, with the exceptions of QED, SA, and Diversity. In \Cref{tab:mol_prop}, \model{} is only surpassed by GraphBP~\cite{liu2022graphbp} in terms of Diversity, yet it exhibits superior performance compared to another diffusion model, TargetDiff~\cite{guan20223d}. According to the average Vina Dock, \model{} generates molecules with high affinities to the pockets (-7.90), which is 6.43\% better than the best autoregressive model baseline, AR-SBDD~\cite{luo20213d}, and 22.16\% better than the best diffusion model baseline, TargetDiff~\cite{guan20223d}. Besides, \model{} surpasses AR-SBDD~\cite{luo20213d} and TargetDiff~\cite{guan20223d} on average High Affinity (60.1\%) by 58.6\% and 6.7\%, and average Diversity (0.74) by 6.71\% and 4.23\%. On the other hand, the SA of generated molecules should fall within a reasonable range so that the ability to explore the molecular space confined by protein pockets is high enough to discover potential molecules. As shown in \Cref{tab:mol_prop}, the average QED of \model{} (0.39) is slightly lower than that of AR-SBDD~\cite{luo20213d} (0.51) and TargetDiff~\cite{guan20223d} (0.48), but remains comparable to that of liGAN~\cite{ragoza2022chemsci} (0.39), implying that \model{} satisfies this desired property. Notably, the molecules generated by \model{} perform even better than those in the test set on Vina Score, Vina Min, and Vina Dock, suggesting that \model{} has great potential to generate more drug-like molecules with higher affinity. \model{} models the constraints between atomic nuclei and their surrounding electron clouds to prevent separation violatio, which is important for the generation of high-affinity and realistically viable pharmaceuticals. Although TargetDiff~\cite{guan20223d}, a diffusion-based model, also generates molecules by sampling from a learned distribution, it solely considers the positional information of atomic nuclei when learning the distribution of atoms, which fails to incorporate fundamental physical laws on electrons. Consequently, it is reasonable to assert that the \model{} model, as a geometric diffusion generative model incorporating the manifold constraints, provides critical insights for the generation of molecules and pharmaceuticals with high binding affinities.
\begin{figure*}[tb]
\centering
\includegraphics[width=\textwidth]{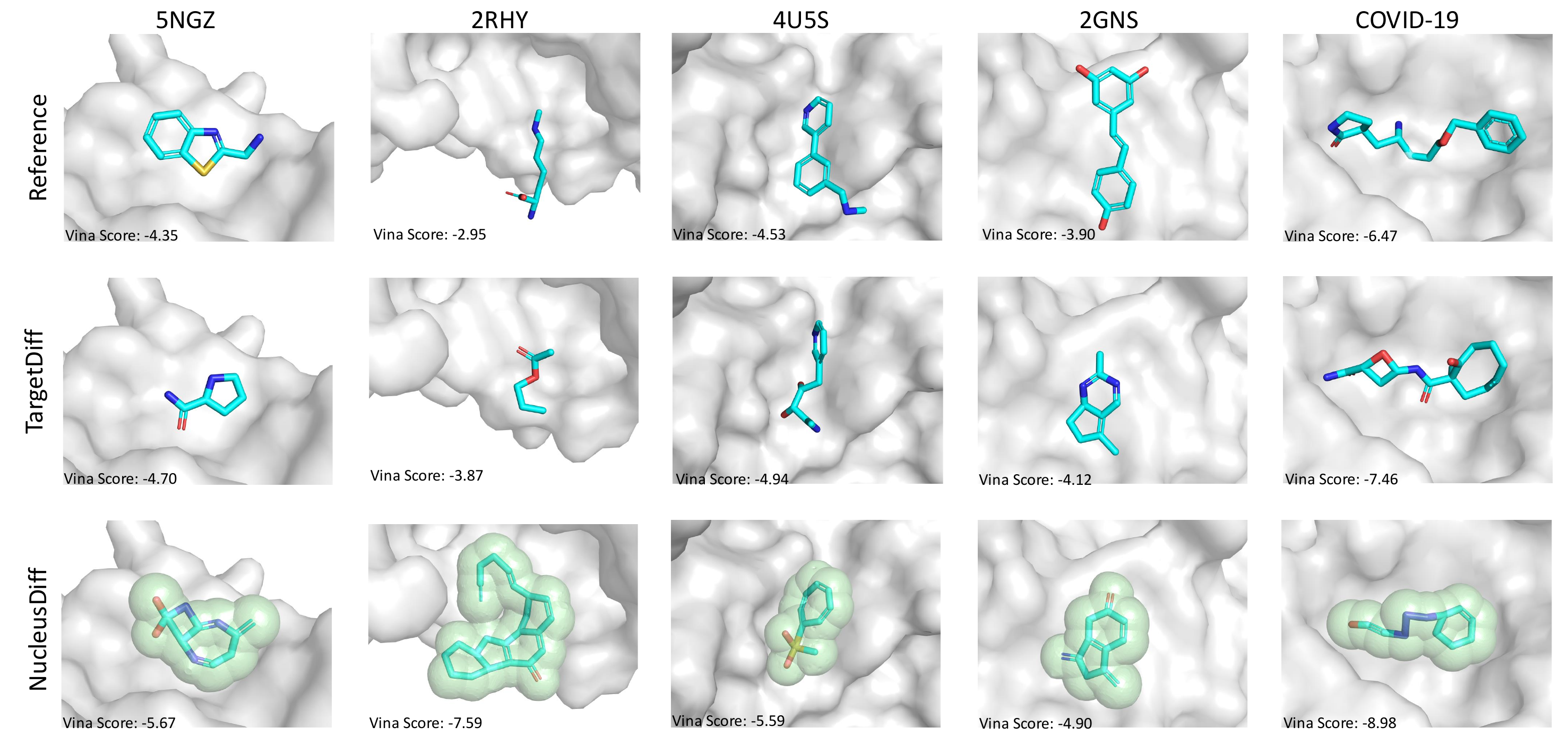}
\vspace{-3ex}
\caption{\small 
Visualization of the pockets and sampled ligands on CrossDocked2020 and COVID-19. The sampled molecules are generated using TargetDiff and \model{}. For \model{}, we illustrate both the generated nucleus and manifold (marked in the green sphere). We also emphasize the use of the Vina Score to measure the quality of generated ligands, where a lower score indicates stronger binding affinity.
}
\label{fig:visualization}
\end{figure*}

\begin{table*}[t!]
\centering
\caption{\small
A summary of 14 biochemical properties for molecules generated by TargetDiff and \model{} for \textbf{target 3CL}. The symbols ($\uparrow$) and ($\downarrow$) indicate a higher or lower value of the metric is preferable, respectively. `Avg.' and `Med.' represent average and median values, respectively.
}
\label{tab2:mol_prop_3cl}
\vspace{-2ex}
\begin{adjustbox}{max width=\textwidth}
\begin{tabular}{c cc cc cc cc cc cc cc}
\toprule[1.5pt]
\multirow{2}{*}{\textbf{Metrics}} & \multicolumn{2}{c}{\textbf{Vina Score ($\downarrow$)}} & \multicolumn{2}{c}{\textbf{Vina Min ($\downarrow$)}} & \multicolumn{2}{c}{\textbf{Vina Dock ($\downarrow$)}} & \multicolumn{2}{c}{\textbf{High Affinity ($\uparrow$)}} & \multicolumn{2}{c}{\textbf{QED ($\uparrow$)}}   & \multicolumn{2}{c}{\textbf{SA ($\uparrow$)}} & \multicolumn{2}{c}{\textbf{Diversity ($\uparrow$)}} \\
\cmidrule(lr){2-3} \cmidrule(lr){4-5} \cmidrule(lr){6-7} \cmidrule(lr){8-9} \cmidrule(lr){10-11} \cmidrule(lr){12-13} \cmidrule(lr){14-15} 
 & \textbf{Avg.} & \textbf{Med.} & \textbf{Avg.} & \textbf{Med.} & \textbf{Avg.} & \textbf{Med.} & \textbf{Avg.} & \textbf{Med.} & \textbf{Avg.} & \textbf{Med.} & \textbf{Avg.} & \textbf{Med.} & \textbf{Avg.} & \textbf{Med.}\\
\midrule
\textbf{TargetDiff} & -4.82   &-5.08  & -5.61  & -5.68 &-6.39  & -6.49 & 50.5\%  &50.5\% & \textbf{0.56}     & \textbf{0.54} &\textbf{0.62}  &   \textbf{0.61}  & \textbf{0.76}    & \textbf{0.76}      \\
\textbf{\model{} (ours)} &\textbf{-5.85}  &\textbf{-5.80}  &\textbf{-6.21}  &\textbf{-6.23}  & \textbf{-6.74} & \textbf{-6.84}  & \textbf{70.0\%} & \textbf{70.0\%}    & 0.43     &0.42  & 0.54 &0.53   & 0.73&0.73  \\
\bottomrule[1.5pt]
\end{tabular}
\end{adjustbox}
\end{table*}

\subsection{Case Study of \model{} on COVID-19 Target}
Scientists and drug designers are particularly interested in how deep learning models generalize to real-world problems, where practical constraints and biological variability should be considered. To this end, we follow previous work and conduct a case study focused on a COVID-19-related therapeutic target~\cite{zhang2023learning}. This case study allows us to assess the practical utility of \model{} in addressing critical and timely challenges in structure-based drug design. In this context, we generate molecules targeting the COVID-19 therapeutic target and subsequently evaluate their binding affinities and drug-likeness properties using the same set of metrics discussed earlier, including the Vina Score, Vina Min, Vina Dock, QED, SA, and Diversity. Lastly, we assess the performance of \model{} and TargetDiff on this COVID-19-related therapeutic target using the proposed PLVR metric.

\paragraph{Separation Violation Evaluation.}
To evaluate the performance of \model{} on the COVID-19 therapeutic target, we conduct a comprehensive comparison with TargetDiff. Both methods utilize diffusion-based models, enabling a thorough understanding of separation violations during the inference process of DDPM (Denoising Diffusion Probabilistic Models)~\cite{ho2020denoising}. Our evaluation involves the 3CL protease as a real-world therapeutic target, with its experimentally validated active ligands. For both \model{} and TargetDiff, we set an identical number of timesteps (1000) for inference. We analyze the violation ratio metric PLVR at 11 timesteps, sampled every 100 steps from Step-0 to Step-1000, and the main results are summarized in~\Cref{tab:collision_plvr}. We observe that TargetDiff shows a significant reduction in separation violations from Step-0 to Step-1000. However, \model{} maintains a consistently lower violation ratio throughout the inference steps. Notably, \model{} outperforms TargetDiff in the PLVR metric, achieving improvements of up to 66.7\%. Specifically, in the final sampling steps, \model{} achieves an almost negligible separation violation ratio, verifying the effectiveness of our design. 

\Cref{fig:atomic_collision_illustration} visualizes the violation ratio for both methods across the COVID-19 target (3CL), revealing a marked contrast in the convergence trends. \model{} demonstrates a more pronounced and rapid reduction in violation ratios compared to TargetDiff, underscoring its potential for addressing challenging targets in real-world drug design, such as those encountered in COVID-19 research.

\paragraph{Binding Affinity Evaluation.}
We compare the performance of \model{} and TargetDiff on a curated benchmark involving the COVID-19 therapeutic target 3CL, assessing the 14 metrics. As shown in \Cref{tab2:mol_prop_3cl}, \model{} outperforms TargetDiff on 8 out of 14 metrics. Specifically, \model{} achieves a superior average Vina Score of -5.85 compared to TargetDiff's -4.82, indicating a stronger binding affinity. \model{} also excels in the Vina Min and Vina Dock metrics, with average scores of -6.21 and -6.74, respectively, compared to -5.61 and -6.39 for TargetDiff. In the context of high-affinity ligands, \model{} generated 70.0\% high-affinity ligands versus 50.5\% for TargetDiff, showing a clear advantage. Although \model{} has a slightly lower average QED score (0.43) compared to TargetDiff (0.56), it still maintains acceptable drug-likeness properties. Additionally, both models exhibit similar performance in synthetic accessibility (SA) and molecular diversity, with \model{} achieving average scores of 0.54 and 0.73, respectively, slightly below TargetDiff’s values of 0.62 and 0.76. These results demonstrate that \model{} is more effective in generating high-affinity ligands while maintaining a balance with other drug design criteria, making it a strong candidate for real-world therapeutic applications, particularly those related to COVID-19. We further assess the generalization capabilities of \model{} and TargetDiff on two additional therapeutic targets, with detailed results and analyses provided in the supplementary materials due to space limitations.

\subsection{The Results for Minimum Distance Constraint}
Our research is dedicated to addressing the critical challenge of separation violations in structure-based drug design. While \model{} incorporates soft constraints during training to mitigate separation violations, an alternative approach involves implementing minimum distance constraints during the sampling process of pre-trained models. In this part, we rigorously evaluate the efficacy of applying minimum distance constraints to the sampling process of pre-trained \model{} and TargetDiff models, assessing the generated molecules' performance in terms of both separation violation performance and binding affinity.

Given that \model{} has demonstrated near-elimination of separation violation on the CrossDocked2020 dataset, we conduct a more representative experiment. This involves examining the properties of 1000 molecules sampled from both \model{} and TargetDiff models using a minimum distance constraint inference process, specifically targeting the COVID-19 target~(3CL). This approach enables a more thorough evaluation of the models' capabilities under minimum distance constraint inference conditions, providing deeper insights into their performance in this context.

\paragraph{Minimum Distance Constraint for the Inference Process of TargetDiff and \model{}.}
The high-level idea of the minimum distance constraint is to correct the distances of atom pairs that exhibit separation violations during the inference process. In this paper, we present two types of minimum distance constraints as post-correction schemes:

\textbf{Minimum Distance Constraint (Parallelogram):} For protein-ligand atom pairs $(a, b)$ with separation violation, we first identify atom $c$ within the ligand that is closest to ligand atom $b$. Clearly, the line connecting the protein-ligand pair in 3D space intersects with a sphere centered at ligand atom $c$, with the covalent radius equal to the distance between $c$ and $b$. One intersection point obviously becomes $b$, while the other intersection point $b'$ becomes the corrected position of $b$ after applying the minimum distance constraint.

\textbf{Minimum Distance Constraint (Circle):} For a protein-ligand pair $(a, b)$ exhibiting separation violation, we first identify the atom $c$ within the ligand that is closest to ligand atom $b$. We ensure that the distance between $a$ and $b$ is less than that between $a$ and $c$. Based on this condition, we construct two spheres: one centered at protein atom $a$ with a radius equal to the covalent radius of $a$ and $b$ (noting that the distance between $(a, b)$ is less than the summation of their covalent radius), and another centered at ligand atom $c$ with a radius equal to the distance between $c$ and $b$. If these two spheres intersect, their intersection forms a circular region. By deriving the analytical expression for this circle, we sample a new point $b'$ on this circle using a predetermined random seed (42). This $b'$ effectively avoids separation violations while preserving the ligand's geometric characteristics. In the case where the two spheres are tangent, the point of tangency serves as the unique corrected position $b'$, avoiding separation violations and maintaining the ligand's geometric properties.

\paragraph{The Separation Violation Evaluation for Minimum Distance Constraint.} The experimental results present in~\cref{table:merged_collision_hard_constraint_plvr} demonstrate the efficacy of implementing minimum distance constraints to mitigate separation violations in structure-based drug design, specifically for the COVID-19 target (3CL). We evaluate the performance using the proposed PLVR metric. For TargetDiff, the baseline model without minimum distance constraints exhibits a non-negligible level of separation violations, with 5 violations per 210,000 atom pairs (PLVR). Notably, the implementation of both parallelogram and circle minimum distance constraints completely eliminates these violations over the PLVR metric, resulting in zero violations for all ratios. Similarly, \model{} shows a slight improvement in the baseline performance compared to TargetDiff, with 3 violations per 210,000 atom pairs (PLVR). This baseline improvement can be attributed to the manifold-constrained modeling approach inherent to \model{}. Nevertheless, the application of minimum distance constraints (both parallelogram and circle methods) yields the same perfect results as observed with TargetDiff, completely eliminating all separation violations.

These results underscore the validity of incorporating minimum distance constraints in the sampling process of pre-trained models for structure-based drug design. Both the parallelogram and circle constraint methods prove equally effective in resolving separation violations, suggesting that either approach can be reliably employed to enhance the physical realism of generated molecular structures. The complete elimination of separation violations over the PLVR metric for both TargetDiff and \model{} when using minimum distance constraints highlights the robustness of this approach. This improvement is particularly significant for the COVID-19 target (3CL), demonstrating the potential of these methods in generating more physically viable drug candidates for this crucial therapeutic target.

\begin{table}[t!]
\caption{\small 
The PLVR performance among pocket-ligand pairs for structure-based drug design of TargetDiff and \model{} in the COVID-19 target. Two types of minimum distance constraints are considered: w/ Parallelogram and w/ Circle.
}
\label{table:merged_collision_hard_constraint_plvr}
\vspace{-10pt}
\centering
\begin{adjustbox}{max width=\textwidth}
\begin{tabular}{l c c}
\toprule[1.5pt]
\multicolumn{1}{l}{\textbf{Models}} & \multicolumn{1}{c}{\textbf{TargetDiff}} & \multicolumn{1}{c}{\textbf{NucleusDiff~(ours)}} \\
\midrule
  - & 5/210000 &  3/210000 \\
  \textbf{+ Parallelogram} & \textbf{0/210000} &  \textbf{0/210000} \\
  \textbf{+ Circle} & \textbf{0/210000} &  \textbf{0/210000} \\
\bottomrule[1.5pt]
\end{tabular}
\end{adjustbox}
\end{table}

\paragraph{The Binding Affinity Evaluation for Minimum Distance Constraint.} 
In this study, we analyze the effects of incorporating a minimum distance constraint to the model's inference process, which influences the physicochemical properties of separation-violating molecules generated by TargetDiff and NucleusDiff for the COVID-19 target, 3CL. We select one representative separation-violating molecule each for TargetDiff and NucleusDiff, with the experimental results detailed in~\cref{table:merged_targetdiff_nucleusdiff}. 

The results prominently demonstrate that enforcing minimum distance constraints often leads to a decrease in binding affinity, a critical factor in drug design. For the separation-violating molecule generated by Targetdiff, this trend is clearly observed. The Vina Score, a key indicator of binding affinity where lower values are more favorable, increases from 19.287 in the unconstrained version to 20.124 with the parallelogram constraint and 20.270 with the circle constraint. This consistent increase in Vina Score across both constraint methods signifies a reduction in binding affinity. The separation-violating molecule generated by \model{} further corroborates this trend, albeit to a lesser extent. While the parallelogram constraint results in an invalid structure, the circle constraint method produces a valid molecule with a slightly higher Vina Score (-5.939) compared to the unconstrained version (-5.946). Although this difference is minimal, it still aligns with the overall trend of decreased binding affinity when constraints are applied. 

These findings underscore a crucial trade-off in the application of minimum distance constraints: while they effectively address separation violations, they often do so at the cost of reduced binding affinity. This phenomenon was consistently observed across both TargetDiff and \model{} methods, indicating that it may be a general consequence of applying such constraints rather than a method-specific effect. The observed decrease in binding affinity highlights the need for careful consideration when applying minimum distance constraints in structure-based drug design. While these constraints serve an important purpose in eliminating separation violations, their potential to compromise binding affinity could have significant implications for drug efficacy.

\begin{table}[t!]
\centering
\caption{\small
The binding affinity performance of separation-violating molecules generated by TargetDiff and \model{} for target 3CL. Two types of minimum distance constraints are considered: w/ Parallelogram and w/ Circle. The symbols ($\uparrow$) and ($\downarrow$) indicate whether a higher or lower value of the metric is preferable, respectively.
}
\label{table:merged_targetdiff_nucleusdiff}
\vspace{-2ex}
\begin{adjustbox}{max width=\textwidth}
\begin{tabular}{l c c c}
\toprule[1.5pt]
\textbf{Method} & \textbf{Vina Score ($\downarrow$)} & \textbf{Vina Min ($\downarrow$)} & \textbf{Vina Dock ($\downarrow$)} \\
\midrule
\textbf{TargetDiff}         & \textbf{19.287}  & \textbf{-0.543}  & \textbf{-6.393}  \\
\textbf{+ Parallelogram}     & 20.124           & -0.363           & -6.387           \\
\textbf{+ Circle}            & 20.270           & -1.827           & -6.075           \\
\midrule
\textbf{\model{} (ours)}    & \textbf{-5.946}  & \textbf{-7.055}  & \textbf{-7.646}  \\
\textbf{+ Parallelogram}     & Invalid          & Invalid          & Invalid          \\
\textbf{+ Circle}            & -5.939           & -6.516           & -6.441           \\
\bottomrule[1.5pt]
\end{tabular}
\end{adjustbox}
\end{table}

\subsection{Visual Analysis of \model{} on CrossDocked2020 and COVID-19 Target}
\Cref{fig:visualization} illustrates the ligands generated by \model{} and TargetDiff, alongside reference ligands, for specific binding pockets. We select five representative protein pockets for structural analysis: 5NGZ, 2RHY, 4U5S, 2GNS from the CrossDocked2020 dataset, and an additional COVID-19 target. Both TargetDiff and \model{} demonstrate the capability to generate structurally diverse ligands that conform to their respective binding pockets. Notably, \model{} consistently produces ligands with superior Vina Scores compared to both TargetDiff and the reference ligands across all examined pockets. This is particularly evident for the COVID-19 target, where \model{} achieves a remarkable Vina Score of -8.98, substantially outperforming TargetDiff (-7.46) and the reference ligand (-6.47). A key distinction between TargetDiff and \model{} lies in the molecular representations they generate. While TargetDiff generates only atomic positions, \model{} provides a more comprehensive visualization by not only displaying atomic nuclei positions but also incorporating the manifold of generated molecules. This is depicted by a green spherical manifold encompassing the ligand structure, providing a deeper understanding of how the molecule occupies space within the binding pocket.

It is important to note that our evaluation of molecular properties focuses on metrics such as binding affinity (Vina score, Vina Min, and Vina Dock), QED (Quantitative Estimate of Drug-likeness), SA (Synthetic Accessibility), and Diversity. Structural similarity between the generated molecule and the reference ligand is not necessarily required, as the generative model is not designed to replicate these structures exactly. Also, the reference ligand is not always the optimal drug candidate for the target, and its structure often leaves room for improvement. As these reference structures are typically derived through computational chemistry methods, the ability of the generative model to produce diverse ligands is a significant advantage.

Due to space limitations, the technical details of the manifold generation process are elaborated in the supplementary materials. However, the manifold representation in \model{} visualizations offers crucial insights: (1) It enables the reconstruction of the ligand's electron cloud, demonstrating that the model has learned to generate structures that adhere to van der Waals radius constraints—a key physical property that is incorporated into our objective function. (2) This approach sets our work apart from previous structure-based drug design efforts by explicitly modeling and visualizing the electron cloud distribution, rather than focusing solely on atomic nuclei positions. The clear delineation of ligand boundaries within protein pockets, especially evident in the manifold representations, suggests that \model{} not only effectively learns the relative positioning of ligands within pockets but also the underlying physical principles governing the distribution of atomic nuclei and electron clouds. This comprehensive approach likely contributes to the generation of ligands with improved binding affinities and a reduced likelihood of violating separation constraints.

\section{Conclusion}
In this work, we first introduce three metrics to effectively quantify the separation violation issue of existing deep generative models for ligand design in structure-based drug discovery tasks. We then present \model{}, which jointly models the atomic nucleus and the surrounding manifold to address this violation issue. Empirical results reveal that \model{} not only achieves superior performance on existing stability and potency metrics but also significantly mitigates separation violations and converges more rapidly to the target geometric distribution.

One limitation of \model{} is its current focus on modeling the ligand manifold while neglecting the protein manifold. Incorporating the protein manifold into the ligand manifold is an important aspect that we aim to explore in future work. Additionally, \model{} utilizes a discretized version of the manifold over the electron cloud, represented by triangle mesh points. Yet in reality, the manifold is continuous. This requires a more profound understanding and utilization of the first-principle theorem, and we would like to leave this for future research.

\section{Materials and Methods}
Our main goal is to jointly model the nucleus and the manifold over the electron cloud surrounding each nucleus, aiming to reduce the separation violation issue in the diffusion model sampling process for structure-based drug design. We first explain how the DDPM model is applied to the existing structure-based drug design modeling. Following this, we introduce how we adopt the manifold-constrained modeling in \model{}. Last but not least, we provide more details on the training objective and inference, along with insights into the architecture specifics.

\subsection{Nucleus Diffusion for Atomic Nuclei Generation}
Here our main goal is to model the nuclei types $\vv^N$ and nuclei coordinates $\vx^N$ given the protein pocket $(\vv^P, \vx^P)$: $p(\vv^N, \vx^N|\vv^P, \vx^P)$. We follow the existing DDPM for the structure-based drug design pipeline by estimating this conditional with a categorical diffusion model on atomic types and a continuous diffusion model on atomic coordinates~\cite{guan20223d,hoogeboom2021argmax}.

In the main manuscript, we define the variance scheduler $\beta_t$ and $\alpha_t$, and how to derive the prior $q(\vx^N_t | \vx^N_0)$ and posterior $q(\vx^N_{t-1} | \vx^N_t, \vx^N_0)$ for nuclei coordinates at time $t$. Similarly, for the nuclei types, we use categorical distribution $C$, and suppose we have $K$ nuclei types in total. The prior distribution and posterior distribution of nuclei types at time $t$ are
\begin{equation}
\begin{aligned}
    q(\vv^N_t | \vv^N_0) &= C(\vv^N_t | \bar \alpha_t \vv^N_0 + (1-\bar \alpha_t) / K),\\
    q(\vv^N_t | \vv^N_t, \vv^N_0) &= C(\vv^N_{t} | \theta_c(\vv^N_t, \vv^N_0)),
\end{aligned}
\end{equation}
where $\theta_c(\vv^N_t, \vv^N_0) = \theta^* / \sum_k \theta^*_k$, and $\theta^* = [\alpha_t \vv^N_t + (1-\alpha_t) / K] \odot [\bar\alpha_{t-1} \vv^N_0 + (1-\bar\alpha_{t-1}) / K]$, where $\odot$ is element-wise product. Then after reparameterization, we predict $\hat \vv^N_0$ from $\vv^N_t$, {\ie}, $\tilde \vv^N_0 = \mu_\theta(\vv^N_t, t)$. Injecting this into the posterior, then the objective functions for the discrete types and continuous coordinates are
\begin{equation} 
\begin{aligned}
\mathcal{L}_{t-1}(\vv^N) & = \text{KL}(q(\vv^N_t | \vv^N_t, \vv^N_0) || q(\vv^N | \tilde \vv^N_0)) \\&= \sum_k \theta_c(\vv^N_t, \vv^N_0)_k \cdot \frac{\theta_c(\vv^N_t, \vv^N_0)_k}{\theta_c(\vv^N_t, \tilde \vv^N_0)_k},\\
\mathcal{L}_{t-1}(\vx^N) & = \mathbb{E}_q \big[ \| \vx^N_0  - \hat \vx^N_0 (\vx^N_t, t, \vv^P, \vx^P) \|^2 \big].
\end{aligned}
\end{equation}

\subsection{Manifold-Constraint Denoising Diffusion Model}
Meanwhile, the generated nuclei coordinates should follow the physical properties: atoms are not solid points and they are composed of nuclei and surrounding electron clouds; thus there exists a minimum distance between pairwise atoms. Ignoring this can lead to the separation violation issue. To alleviate this, we jointly model the manifold and nuclei for structure-based drug design.

To be more concrete, for each nucleus, we construct a discrete manifold, where the radius is the van der Waals radius $R$. Then for each nucleus, we obtain its $c$ closest mesh points in the manifold, marked as $\vx^M$. Thus, instead of $p(\vv^N, \vx^N|\vv^P, \vx^P)$, the objective is to maximize the following likelihood $p(\vv^N, \vx^N, \vx^M|\vv^P, \vx^P)$. The objective on the manifold at time $t$ is \Cref{eq:kl_divergence_in_DDPM_mesh_points}.

On the other hand, recall that the mesh points are scattered around the nuclei with van der Waals radius $R$. Motivated by this, we add a regularization term by forcing the distance between each mesh point and nuclei to be close to $R$ as in~\Cref{eq:regularization_term}.

\subsection{Learning and Inference}
To sum up, the training objective function is composed of four parts, as in~\Cref{eq:overall_objective}. For the inference, because the mesh points from manifold modeling are treated as the auxiliary components of the physics-guided nuclei modeling, they can be ignored at this stage, while only the nuclei coordinates are considered. We provide the detailed pseudo-codes for inference in the supplementary materials.

\subsection{Computational Resources}
All algorithms and models have been developed using Python 3.8.13, with PyTorch version 1.12.1 and PyTorch Geometric version 2.5.2, under CUDA 11.0. Experiments are conducted on a server with 8 NVIDIA V100 GPUs (32 GB memory) and Intel(R) Xeon (R) Platinum 8255C CPU @ 2.50GHz. We employ a single V100 GPU for training while leveraging eight GPUs to accelerate the sampling procedure. The models typically converge after approximately 48 hours of training and sampling 10k ligands using eight GPUs takes about 12 hours. 

\section*{Data, Materials, and Software Availability}
We provide the code and dataset generation scripts at this \href{https://github.com/yanliang3612/NucleusDiff}{GitHub repository}. The code developed for this study is released under the MIT License.

\section*{Author Contributions Statement}
S.L., D.Y., W.D., W.L., Z.L., H.G., C.B., J.C., and A.A. conceived and designed the experiments.
S.L. and D.Y. analyzed the data.
D.Y. performed the experiments.
S.L., D.Y., W.L., and Z.L. contributed to the visual analysis and demos.
S.L., D.Y., W.D., Z.L., and W.L. contributed analysis tools.
S.L., D.Y., W.D., H.G., C.B., J.C., and A.A. wrote the paper.
C.B., J.C., and A.A. contributed equally to advising this project.

\section*{Competing Interests Statement}
The authors declare no competing interests.

\bibliography{reference_article}

\newpage
\appendix

\input{appendix/01_related_work}

\clearpage
\input{appendix/02_collision_metrics}

\clearpage
\input{appendix/03_manifold_construction}

\clearpage
\input{appendix/04_model_details}

\clearpage
\input{appendix/05_exp_setups}

\clearpage
\input{appendix/06_exp_results}

\clearpage
\input{appendix/07_ablation_studies}

\end{document}

%% file: math_commands.tex
\usepackage{amsmath,amsfonts,bm}

\def\eqref#1{equation~\ref{#1}}

\def\1{\bm{1}}

\def\vf{{\bm{f}}}

\def\vh{{\bm{h}}}

\def\vv{{\bm{v}}}

\def\vx{{\bm{x}}}

\DeclareMathAlphabet{\mathsfit}{\encodingdefault}{\sfdefault}{m}{sl}
\SetMathAlphabet{\mathsfit}{bold}{\encodingdefault}{\sfdefault}{bx}{n}

\def\gM{{\mathcal{M}}}

\def\gP{{\mathcal{P}}}



%% file: appendix/01_related_work.tex
\section{Related Work}
\paragraph{Structure-based Drug Design.} 
In recent years, the availability of structural data catalyzes the development of numerous generative models for the \emph{target-aware} molecule generation task. Such models include those by~\cite{skalic2019target, xu2021novo}, which generate SMILES representations based on protein contexts, and the flow-based model proposed by~\cite{tan2022target} for generating molecular graphs conditional on protein target sequence embeddings. \cite{ragoza2022chemsci}~explores the generation of 3D molecules through the voxelization of molecules in atomic density grids within a conditional VAE framework. \cite{sun2023graphvfcontrollableproteinspecific3d}~leverages normalizing flow to generate pocket-based 3D molecules, with properties controlled by a tailored prior distribution. Further, \cite{li2021structure} employs Monte-Carlo Tree Search coupled with a policy network for optimizing molecules in 3D space. Notably, \cite{luo20213d, liu2022graphbp, peng2022pocket2mol}~develop autoregressive models for atom-by-atom 3D molecule generation using Graph Neural Networks (GNNs). Despite these advancements, current models still grapple with several challenges. These include the separate encoding of small molecules and protein pockets~\cite{skalic2019target, xu2021novo, tan2022target, ragoza2022chemsci}, reliance on voxelization techniques and non-equivariant networks~\cite{skalic2019target, xu2021novo, ragoza2022chemsci}, and the limitations inherent to autoregressive sampling methods~\cite{luo20213d, liu2022graphbp, peng2022pocket2mol}. In contrast to previous approaches, our equivariant diffusion-based generative model innovatively integrates 3D protein-ligand interactions within a unified framework while employing non-autoregressive sampling, thereby enhancing the model's capacity to capture complex molecular relationships and improving the consistency between training and inference processes.

\paragraph{Molecular and Protein Manifold Learning.} 
Manifold learning has been widely applied in the vision tasks~\cite{lin2008riemannian,he2003locality,tenenbaum2000global,roweis2000nonlinear,liu2017sphereface}.
Additionally, recent advancements in manifold learning for molecular \& protein have garnered attention in the scientific community. Several studies~\cite{gainza2020deciphering, zhang2023learning, wang2022learning, mallet2023atomsurf} embark on an innovative trajectory by employing or integrating sophisticated representational learning techniques pertaining to molecular and protein surfaces. This approach facilitates a precise articulation and comprehension of the intricate complexities inherent in molecular structures.\cite{gainza2020deciphering} introduces MaSIF, a method using deep learning to identify and predict how proteins interact with other molecules by analyzing patterns on their surfaces. SurfGen~\cite{zhang2023learning}, introducing its two neural networks, Geodesic-GNN and Geoatom-GNN, effectively analyzes topological interactions on pocket surfaces and spatial interactions between ligand atoms and surface nodes for advanced molecular prediction. HMR~\cite{wang2022learning} employs Laplace-Beltrami eigenfunctions for representing molecules on 2D Riemannian manifolds, enhancing molecular encoding through harmonic message passing. Atomsurf~\cite{mallet2023atomsurf} explores the use of 3D mesh surfaces for representing proteins, revealing that while promising, this method alone is less effective than 3D grids. It proposes a novel framework that synergistically combines surface representations with graph-based methods for improved protein representation learning.

%% file: appendix/02_collision_metrics.tex
\section{Formulation and Analysis of Proposed Separation Violation Metrics} \label{sec:SI:collision_metrics}

The phenomenon of separation violation arises when two atoms come sufficiently close to each other such that their electron clouds overlap. This overlap violates the Pauli exclusion principle and the electrostatic repulsion between electrons. To quantify this proximity, we employ covalent radius, denoted as $d$. Consider a scenario involving a ligand atom with coordinates $\vx_i$ and a protein atom with coordinates $\vx_j$. The respective covalent radii for these atoms are $d_i$ and $d_j$. During the ligand generation sampling process, if the distance between these two atoms becomes excessively small, specifically if $\|\vx_i - \vx_j \| \leq D_{ij} = d_i + d_j$, we classify this event as an \textit{separation violation}. To systematically analyze and understand this phenomenon, we propose three distinct metrics for its quantification, each derived from different levels of assessment granularity. These metrics provide a comprehensive framework for evaluating the occurrence and implications of separation violations in molecular simulations and structural biology studies.

\paragraph{Pairwise-Level Violation Ratio (PLVR).}

The first metric we introduce is the atom \textbf{Pairwise-Level Violation Ratio (PLVR)}. This metric quantifies the violation ratio between ligand atoms and their nearby protein atoms within the binding site. The PLVR provides a detailed measure of how often separation violations occur at the pairwise level between ligand and protein atoms within a binding site. For each ligand atom, denoted as $\vx^L_i$, we identify its $K$ nearest protein atoms within the binding site. This step ensures that we consider only the most relevant interactions in the context of the ligand-protein binding interface. The PLVR is then defined mathematically as follows:

\begin{equation}
\begin{aligned}
    \text{PLVR} = \frac{\sum_{k \in N_{\text{mol}}, i \in N^k_{\text{atom}}, j \in N^i_{\text{nearest}}} \mathbbm{1} (\| \vx_i - \vx_j \| < D_{ij})}{K \cdot \sum_{k \in N_{\text{mol}}}N^k_{\text{atom}} },
\end{aligned}
\end{equation}

where the variables are defined as:
\begin{itemize}
    \item $N_{\text{mol}}$: The total number of ligand molecules.
    \item $N^k_{\text{atom}}$: The number of atoms in the $k$-th ligand molecule.
    \item $N^i_{\text{nearest}}$: The number of the nearest protein atoms to the $i$-th ligand atom.
    \item $\mathbbm{1}(\cdot)$: The indicator function, which returns 1 if the condition inside is true (i.e., if the distance between the $i$-th ligand atom and the $j$-th protein atom is less than $D_{ij}$), and 0 otherwise.
    \item $\| \vx_i - \vx_j \|$: The Euclidean distance between the coordinates of the $i$-th ligand atom and the $j$-th protein atom.
    \item $D_{ij}$: The sum of the covalent radius of the $i$-th ligand atom and the $j$-th protein atom.
\end{itemize}

The PLVR metric provides a normalized measure of the frequency of separation violations, which allows for a comparative analysis across different ligand-protein systems. This metric is particularly useful for assessing the spatial compatibility of ligands within protein binding sites, and it can be used to guide the optimization of ligand design to minimize unfavorable atomic interactions. The PLVR metric provides a rigorous and systematic evaluation of separation violations. This facilitates more accurate and informative assessments in molecular simulations and structural biology research.

\paragraph{Atom-Level Violation Ratio (ALVR).}
The second metric we propose is the \textbf{Atom-Level Violation Ratio (ALVR)}. This metric provides a more aggregated view by focusing on individual ligand atoms and determining whether they are involved in any violations with protein atoms. The ALVR measures the proportion of ligand atoms that have at least one of their $K$ nearest protein atoms within a critical distance. This critical distance is defined as less than the sum of their respective covalent radii.

More rigorously, an atom is flagged as involved in a violation if any one of its $K$ nearest neighboring protein atoms is within this critical distance threshold. The ALVR  provides a broader perspective on the violation propensity of each ligand atom, moving beyond pairwise interactions to a more holistic assessment.

The ALVR is mathematically defined as follows:
\begin{equation}
\begin{aligned}
    \text{ALVR} = \frac{ \sum_{k \in N_{\text{mol}}, i \in N^k_{\text{atom}}} \mathbbm{1}\left(\sum_{j \in N_{\text{nearest}}^{i}} \mathbbm{1} (\| \vx_i - \vx_j \| < D_{ij}) > 0\right)}{\sum_{k \in N_{\text{mol}}}N^k_{\text{atom}}},
\end{aligned}
\end{equation}

where the variables are defined as:
\begin{itemize}
    \item $N_{\text{mol}}$: The total number of ligand molecules.
    \item $N^k_{\text{atom}}$: The number of atoms in the $k$-th ligand molecule.
    \item $N^i_{\text{nearest}}$: The number of the nearest protein atoms to the $i$-th ligand atom.
    \item $\mathbbm{1}(\cdot)$: The indicator function, which returns 1 if the condition inside is true and 0 otherwise.
    \item $\| \vx_i - \vx_j \|$: The Euclidean distance between the coordinates of the $i$-th ligand atom and the $j$-th protein atom.
    \item $D_{ij}$: The sum of the covalent radius of the $i$-th ligand atom and the $j$-th protein atom.
\end{itemize}

In this definition, the inner $\mathbbm{1}(\cdot)$ function checks if the distance between the $i$-th ligand atom and any of its $K$ nearest protein atoms is smaller than $D_{ij}$. If at least one such pair is found, the outer $\mathbbm{1}(\cdot)$ function marks the $i$-th ligand atom as involved in a violation.

The ALVR provides a normalized measure of the frequency of separation violations at the atom level. This enables researchers to identify specific ligand atoms that are more prone to violations. The metric is particularly valuable for guiding ligand optimization processes, as it ensures that individual atomic interactions are considered in the context of the entire ligand-protein binding interface. The ALVR enables researchers to achieve a more nuanced understanding of violation dynamics, ultimately facilitating the design of more compatible and effective ligands in molecular simulations and structural biology studies.

\paragraph{Molecule-Level Violation Ratio (MLVR).}

The final metric we introduce is the \textbf{Molecule-Level Violation Ratio (MLVR)}. This metric applies the violation analysis to the entire molecule, providing a macroscopic view of violation occurrences. The MLVR quantifies the proportion of ligand molecules that exhibit at least one atomic violation within their structure. In essence, a molecule is marked as involved in a violation if any of its constituent atoms is flagged for an atom-level violation.

To compute the MLVR, we aggregate the atom-level violations for each molecule. If any atom within a molecule meets the violation criteria (i.e., it has at least one of its $K$ nearest protein atoms within the critical distance threshold), the entire molecule is considered to be in violation.

The MLVR is mathematically defined as follows:
\begin{equation}
\begin{aligned}
    \text{MLVR} = \frac{\sum_{k \in N_{\text{mol}}} \mathbbm{1}\left(\sum_{i \in N^k_{\text{atom}}, j \in N_{\text{nearest}}^{i}} \mathbbm{1} (\| \vx_i - \vx_j \| < D_{ij}) > 0\right)}{N_{\text{mol}}},
\end{aligned}
\end{equation}
where the variables are defined as:

\begin{itemize}
    \item $N_{\text{mol}}$: The total number of ligand molecules.
    \item $N^k_{\text{atom}}$: The number of atoms in the $k$-th ligand molecule.
    \item $N^i_{\text{nearest}}$: The number of the nearest protein atoms to the $i$-th ligand atom.
    \item $\mathbbm{1}(\cdot)$: The indicator function, which returns 1 if the condition inside is true and 0 otherwise.
    \item $\| \vx_i - \vx_j \|$: The Euclidean distance between the coordinates of the $i$-th ligand atom and the $j$-th protein atom.
    \item $D_{ij}$: The sum of the covalent radius of the $i$-th ligand atom and the $j$-th protein atom.
\end{itemize}

In this context, the inner $\mathbbm{1}(\cdot)$ function checks if the distance between any pair of the $i$-th ligand atom and its $K$ nearest protein atoms is smaller than $D_{ij}$. If at least one such pair is found, the outer $\mathbbm{1}(\cdot)$ function flags the molecule as having a violation.

The MLVR provides a normalized measure of the frequency of separation violations at the molecule level. This metric captures the overall violation propensity of entire ligand molecules and is particularly valuable for identifying and optimizing ligand molecules with minimal violation tendencies. Consequently, it ensures better compatibility and stability when ligands bind to protein targets. The MLVR enables the researchers to achieve a comprehensive understanding of violation dynamics across different ligand molecules, thereby facilitating the design and selection of more effective and structurally compatible ligands in molecular design and structural biology studies.

\paragraph{Summary.}
The three defined metrics, PLVR (Pairwise-Level Violation Ratio), ALVR (Atom-Level Violation Ratio), and MLVR (Molecule-Level Violation Ratio), serve as pivotal indicators for assessing separation violation between protein pockets and generated ligands. These metrics elucidate the machine learning inference process within the context of Structure-Based Drug Design (SBDD) tasks. Each metric provides a unique perspective: PLVR evaluates the ratio of violative pairs to total atom pairs, ALVR examines the ratio of violative atoms to total atoms, and MLVR assesses the ratio of violative molecules to total molecules.

%% file: appendix/03_manifold_construction.tex
\section{Details for Manifold Reconstruction}

In this part, we provide a detailed description of the technical process employed for the manifold reconstruction from mesh points, as applied in this work. The reconstruction process includes several critical steps, as follows::

\paragraph{Identifying Proximal Mesh Points for Each Atom.} The primary objective is to demonstrate the manifold constructed from the mesh points generated and sampled by our pre-trained model. To achieve this, we identify all mesh points that are in close proximity to each atom. By calculating the distance, we can obtain mesh points whose distances to the sampled atom are close to the van der Waals radius.

\paragraph{Manifold Reconstruction for Each Individual Atom.} Using the \texttt{MeshLab} software, we calculate the mesh for each individual atom. Specifically, we employ the \texttt{Poisson reconstruction} function in \texttt{MeshLab} , which is crucial for generating a smooth and continuous surface from the scattered mesh points. This step is essential for visualizing the atomic structure and understanding the distribution of mesh points around each atom.

\paragraph{Addressing the Sampling Problem.} A significant challenge is that the number of reasonable mesh points obtained from a single sampling is typically too small, resulting in unreasonable meshes for each atom. This issue requires a solution for effective manifold reconstruction. We address this problem using two main strategies. First, we employ multiple sampling to obtain a sufficient number of reasonable mesh points. These points are then combined into a single manifold. Different random seeds ensure variability and robustness in the sampling process. Second, we utilize data augmentation techniques to artificially increase the diversity and quantity of the data. These techniques include transformations such as scaling, rotation, and translation of the mesh points. Both strategies work in tandem to enhance the robustness of the manifold reconstruction, providing a more comprehensive and accurate representation of the atomic structure.

\paragraph{Fusing Different Meshes Using \texttt{TriMesh} Software.} We utilize the \texttt{TriMesh} software to fuse the different meshes obtained from multiple sampling. The process involves loading the mesh files, filling any holes, updating faces to remove duplicates and degenerate faces, and validating the resultant mesh. \texttt{TriMesh} is particularly effective for handling complex mesh operations and ensuring that the final mesh is a closed volume.

\paragraph{Smoothing the Mesh and Final Reconstruction.} Some areas of the mesh are too abrupt and require smoothing. We apply Laplacian smoothing in \texttt{MeshLab} to clean up the manifold, ensuring a more natural and accurate representation. This smoothing process helps eliminate artifacts and irregularities, resulting in a cleaner and more visually appealing mesh. The final reconstructed manifold effectively demonstrates the success of our manifold reconstruction process, illustrating both the atomic nuclei and the electron cloud. This comprehensive approach ensures a detailed and accurate representation of the atomic structure. It provides valuable insights into the spatial distribution and interactions within the molecule.

%% file: appendix/04_model_details.tex
\section{\model{}'s Details}

\subsection{Difference Between Manifold Constraint and Separation Violation Constraint In Learning}

\begin{figure}[ht]
\centering
\includegraphics[width=.9\textwidth]{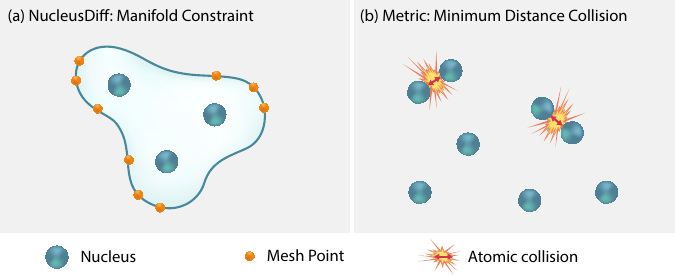}
\vspace{-2ex}
\caption{The comparison of using manifold constraint and minimum distance constraint.}
\label{fig:appendix:comparison_of_manifold_constraint_and_atomic_collision_constraint}
\end{figure}

We have explained how the three violation metrics are constructed in the main article and~\Cref{sec:SI:collision_metrics}. A simple approach is to use these three metrics as regularization terms during modeling. However, we emphasize two main differences between using manifold constraints and separation violation metrics as constraints, which is also visualized in~\Cref{fig:appendix:comparison_of_manifold_constraint_and_atomic_collision_constraint}.

\begin{itemize}[noitemsep,topsep=0pt]
    \item \textbf{Physical Meaning.} The separation violation metrics discussed so far are atom-wise, thus they can be viewed as from the classic physics aspect. However, the intuition behind the manifold constraint is the manifold learning on the electron clouds, which is quantum physics. We highlight that from this aspect, \model{} using manifold-constraint opens a novel research paradigm by incorporating quantum physics for structure-based drug design.
    
    \item \textbf{Efficiency.} Suppose we have $N$ atoms, then with the separation violation constraint, we must ensure the minimum distance for each atom pair, where the complexity is $O(N^2)$. However, for the manifold constraint, we only need to guarantee the minimum distance between each atom and its electron cloud manifold, {\ie}, the complexity if $O(NM)$ where $M$ is the number of sampled mesh points on the manifold. In our experiments, we take $M=3$, thus using manifold constraint in \model{} is more computationally efficient than using separation violation constraint in learning.
    
\end{itemize}

We also note that we can add the separation violation constraint during inference, yet this leads to worse binding affinity performance. Please check~\Cref{sec:SI:ablation_studies} for more details and results.

\subsection{Architecture Details}

In this subsection, we introduce details on the model architecture. The modeling on 3D atom positions needs to satisfy two important physical properties: such modeling should be equivariant to rotation and translation of the whole molecule system, {\ie}, SE(3)-equivariance. Inspired by the recent advancements in equivariant neural networks~\cite{satorras2021n,liu2023symmetry,du2023molecule,liu2023group,liu2022molecular}, we use a SE(3)-Equivariant Graph Neural Network~\cite{satorras2021egnn} to model the interactions between atoms of the ligand  and the protein pocket, as well as the manifold of ligand: 
\begin{equation}
    [\hat{x}_0, \hat{h}_0] = \phi_\theta(\mathcal{M}_t, t, \mathcal{P}) = \phi_\theta([x_t, h_t], t, \mathcal{P}) 
    ,
\end{equation}
and
\begin{equation}
    [\hat{v}_0, \hat{h}_0] = \phi_\theta(\mathcal{M}_t, t, \mathcal{P}) = \phi_\theta([v_t, h_t], t, \mathcal{P}) 
    .
\end{equation}

In the $l$-th layer, we alternately update the atom's hidden embedding, denoted as $h$, and its coordinates, represented by $x$, in the following manner:
\begin{equation} \label{eq:layer_egnn}
\begin{aligned}
    h^{l+1}_i &= h^{l}_i + \sum_{j\in{\mathcal{V}}, i\neq j}f_h(d^l_{ij}, h^l_i, h^l_j, e_{ij}; \theta_h) &  \\
    x^{l+1}_i &= x^{l}_i 
    + \sum_{j\in{\mathcal{V}}, i\neq j}(x^l_i-x^l_j) f_x(d^l_{ij}, h^{l+1}_i, h^{l+1}_j, e_{ij};
    \theta_x) \cdot 
    \mathds{1}_{\text{mol}}
\end{aligned}
\end{equation}

We consider the distance $d_{ij}$ as the Euclidean distance, which measures the straight line distance between two atoms, atom $i$ and atom $j$. Additionally, $\mathbf{e}_{ij}$ is a feature that identifies whether the connection is between two protein atoms, two ligand atoms, or one of each. We also use a ligand molecule mask, denoted as $\mathds{1}_{\text{mol}}$, to ensure that we only adjust the coordinates of ligand atoms, not protein atoms.

Initially, each atom is represented by a hidden embedding, $\mathbf{h}^0$, created by an embedding layer that incorporates atom-specific information. After processing through several layers, we obtain the final atom hidden embedding, $\mathbf{h}^L$. This embedding is then processed through a multi-layer perceptron and a softmax function to predict $\hat{\mathbf{v}}_0$.

An important aspect of this process ensures rotational equivariance. This means that if we rotate the initial atomic coordinates $\mathbf{x}_t$, the predicted coordinates $\hat{\mathbf{x}}_0$ will rotate in the same way. This is also true for the relationship between $\mathbf{x}_{t-1}$ and $\mathbf{x}_{0}$.

In summary, this process involves calculating distances between atoms, identifying atom types, and transforming atom information through several layers to predict new coordinates, all while maintaining consistency under rotation.


\newpage
\subsection{Algorithm}
\label{appen:algorithm}
In this part, we present a comprehensive overview of the training and sampling methodologies employed by NucleusDiff, which are delineated as Algorithm \ref{alg:training} for training and Algorithm \ref{alg:sampling} for sampling.
\begin{algorithm}[htbp]
    \footnotesize
    \caption{Training  algorithm of NucleusDiff}
    \label{alg:training}
    {
	\begin{algorithmic}[1]
			\STATE {\bfseries Input:} Protein-ligand binding dataset $\{\mathcal{P}, \mathcal{M}\}_{i=1}^N$, Ligand Manifold$\{\mathcal{V}_{i=1}^K, \mathcal{F}_{i=1}^L\}$, neural network $\phi_\theta$ for modeling ligand distribution (parameterized by $\theta$), $\psi_\theta$ for modeling mesh points distribution.
                \WHILE{$\phi_\theta$ not converge}
                 \STATE Identify the $n$ closest mesh points for each atom $\{\mathcal{V}_{j=1}^n\}_{i=1}^N$, thereby obtaining the final dataset $\{\mathcal{P}, \mathcal{M}, \mathcal{V}_{j=1}^n\}_{i=1}^N$.
                    
			  \STATE Move the complex~(including the mesh points) to make Center of Mass (CoM) of protein atoms zero.
       
                \STATE Perturb ligand postion $\vx_0$ to obtain $\vx_t$: $\vx_t = \sqrt{\bar{\alpha}_t} \vx_0 + (1 - \bar{\alpha}_t)\epsilon$, where $\epsilon \in N(0, \bm{I})$.
                
                 \FOR{$ \text{the mesh points' postions}~\vv^i$ in $\{\mathcal{V}_{i=1}^n\}$}
                 
                     \STATE Perturb $\vv^i_0$ to obtain $\vv^i_t$: $\vv^i_t = \sqrt{\bar{\alpha}_t} \vv^i_0 + (1 - \bar{\alpha}_t)\epsilon$, where $\epsilon \in N(0, \bm{I})$,
                     
                 \ENDFOR
                 
                \STATE Perturb $\vf_0$ to obtain $\vf_t$: 
                
                 $log c = \log \left(\bar\alpha_t \vf_0 + (1 - \bar\alpha_t) / K\right)$,
                
                 $\vf_t = \texttt{one\_hot}(\text{argmax}_i [g_i + log c_{i}])$, where  $g \sim \text{Gumbel}(0, 1)$,

                \STATE Uniformly encode the mesh points as $\vh_0$ to distinguish between different atom types.
                
                \STATE Perturb $\vh_0$ to obtain $\vh_t$: 
                
                 $log c' = \log \left(\bar\alpha_t \vh_0 + (1 - \bar\alpha_t) / K\right)$,
                
                 $\vh_t = \texttt{one\_hot}(\text{argmax}_i [g_i + log c_{i}])$, where  $g \sim \text{Gumbel}(0, 1)$,
                \STATE Predict $[\hat \vx_0, \hat \vf_0]$ from $[\vx_t, \vf_t]$ with $\phi_\theta$: $[\hat \vx_0, \hat \vf_0] = \phi_\theta([\vx_t, \vf_t], t, P)$.
                
                \FOR{$\vv^i$ in $\{\mathcal{V}_{i=1}^n\}$}
                \STATE Predict $[\hat \vv^i_0, \hat \vh_0]$ from $[\vv^i_t, \vh_t]$ with $\psi_\theta$: $[\hat \vv^i_0, \hat \vh_0] = \phi_\theta([\vv^i_t, \vh_t], t, P)$,
                \ENDFOR

                \STATE Compute the posterior atom types $\bm{c}(\vf_t, \vf_0)$ and $\bm{c}(\vf_t, \hat{\vf_0})$. 
                \STATE Compute the posterior atom types $\bm{c}(\vh_t, \vh_0)$ and $\bm{c}(\vh_t, \hat{\vh_0})$.
                
                \STATE Compute the unweighted MSE loss on atom coordinates, the KL loss on posterior atom types and weighted manifold-constrained loss: $L = \|\vx_0 - \hat{\vx_0} \|^2 + \text{ KL}(\bm{c}(\vf_t, \vf_0) \parallel \bm{c}(\vf_t, \hat{\vf_0}))$ + $\|\vv_0 - \hat{\vv_0} \|^2$ + $\text{ KL}(\bm{c}(\vh_t, \vh_0) \parallel \bm{c}(\vh_t, \hat{\vh_0}))$ + $\sum_{i=1}^n \| \| \hat \vv^i_0 - \hat{\vx_0} \| - d_{\text{van der Waals}} \|$.
                \ENDWHILE
	\end{algorithmic}}
\end{algorithm}

\begin{algorithm}[htbp]
\caption{Sampling  algorithm of NucleusDiff}
    \label{alg:sampling}
    {
        \begin{algorithmic}[1]
            \STATE {\bfseries Input:} The protein binding site $\gP$, the learned model $\phi_\theta$ for modeling ligand distribution.
            
            \STATE Sample the number of atoms in $\gM$ based on a prior distribution conditioned on the pocket size.
            \STATE Move CoM of protein atoms to zero.
            \STATE Sample initial molecular atom coordinates $\vx_T$ and atom types $\vv_T$: 
            
            $\vx_T \in N(0, \bm{I})$,
            
            $\vf_T = \texttt{one\_hot}(\text{argmax}_i g_i)$, where $g \sim \text{Gumbel}(0, 1)$,
            
            \FOR {$t$ in $T, T-1, \dots, 1$}
                \STATE Predict $[\hat{\vx_0}, \hat{\vf_0}]$ from $[\vx_t, \vf_t]$ with $\phi_\theta$: $[\hat{\vx_0}, \hat{\vf_0}] = \phi_\theta([\vx_t, \vf_t], t, P)$,
                \STATE Sample $\vx_{t-1}$ from the posterior $p_\theta(\vx_{t-1} | \vx_{t}, \hat{\vx_{0}})$,
                \STATE Sample $\vf_{t-1}$ from the posterior $p_\theta(\vf_{t-1} | \vf_{t}, \hat{\vf_{0}})$.
            \ENDFOR
       \end{algorithmic}}
\end{algorithm}

%% file: appendix/05_exp_setups.tex
\section{Experiment Setup}
In this section, we present our approach for constructing CrossDock datasets, describe our evaluation protocol, and provide comprehensive details on our algorithm as well as the baseline methods utilized in our study. To achieve this, we leverage sophisticated techniques and utilize advanced metrics to ensure the reliability and relevance of our results.

\subsection{CrossDock Datasets}

\paragraph{CrossDock Datasets Construction.}
We conduct experiments to evaluate the generative performance of NucleusDiff on the CrossDocked2020 dataset~\cite{francoeur2020three}. This dataset comprises 22.5 million docked protein-ligand pairs, with each pair exhibiting various poses across multiple pockets within the Protein Data Bank. The ligands associated with specific pockets were docked with each receptor assigned to those pockets using smina through Pocketome. Binding data (pK) for the CrossDocked2020 set were sourced from PDBbind v2017, revealing that 41.9\% of the complexes have available binding affinity data. For a fair comparison, we follow previous works~\cite{luo2022autoregressive, guan20223d} by selecting only binding pose data with root-mean-squared deviations (RMSD) of less than 1 \AA\ . we further refine the dataset through clustering at 30\% sequence identity using~\texttt{MMseqs2}~\cite{steinegger2017mmseqs2}. This process yields 100,000 pairs for training and 100 pairs for evaluation.

\paragraph{CrossDock Mesh Datasets Construction.} We utilize \texttt{MSMS}~\cite{ewing2010msms} to compute the solvent-excluded surface of the molecule, employing a probe radius of 1.5\,\AA\ and a sampling density of 3.0 for small molecules, generating a triangular mesh representation. To further refine the surface mesh, we employ \texttt{PyMesh}~\cite{zhou2019pymesh}, which helps in reducing the number of vertices and correcting poorly meshed regions. Addressing degenerate vertices or disconnected surfaces is crucial, as these issues can lead to an improper distribution of mesh points when training the models. Finally, we selected the $K$ mesh points that are closest to the van der Waals radii distance from the nucleus to construct a mesh point dataset for the ligand. This dataset predominantly includes the 3D coordinates of the mesh points.

\subsection{Training Details}
The two score model are trained using the gradient descent method Adam~\cite{kingma2014adam} with \texttt{init\_learning\_rate=0.001}, \texttt{betas=(0.95, 0.999)}, \texttt{batch\_size=4}, and \texttt{clip\_gradient\_norm=8}. To balance the scales of the two losses, we apply a factor of $\alpha=100$ to the atom type loss. During the training phase, we add small Gaussian noise with a standard deviation of 0.1 to protein atom coordinates as data augmentation. We also schedule to decay the learning rate exponentially with \texttt{a factor of 0.6} and \texttt{a minimum learning rate of 1e-6}. The learning rate is decayed if there is no improvement in the validation loss over \texttt{10 consecutive evaluations}. The evaluation is performed for \texttt{every 100 training steps}.

\subsection{Implementation Details}
Our NucleusDiff comprises two score models, each consisting of 9 equivariant layers. Each layer is a Transformer~\cite{vaswani2017attention} with \texttt{hidden\_dim=128} and \texttt{n\_heads=16}. The key/value embeddings and attention scores are generated through a 2-layer MLP with LayerNorm and ReLU activation. For atom coordinates, we use a sigmoid $\beta$ schedule with $\beta_1 = \texttt{1e-7}$ and $\beta_T = \texttt{2e-3}$. For atom types, we adopt a cosine $\beta$ schedule as suggested by \cite{nichol2021improved}, with \texttt{s=0.01}. We set the number of diffusion timesteps to \texttt{1000}.

\subsection{Baselines}
We conducted a comparative evaluation of NucleusDiff with leading generative models for structure-based drug design, including liGAN~\footnoteE{\small LiGAN (GPL-2.0 license): \url{https://github.com/mattragoza/LiGAN}.}, GraphBP~\footnoteE{\small GraphBP (GPL-3.0 license): \url{https://github.com/divelab/GraphBP}.}, AR-SBDD~\footnoteE{\small AR-SBDD (MIT license): \url{https://github.com/luost26/3D-Generative-SBDD}.}, Pocket2Mol~\footnoteE{\small Pocket2Mol (MIT license): \url{https://github.com/pengxingang/Pocket2Mol}.}, and TargetDiff~\footnoteE{\small TargetDiff (MIT license): \url{https://github.com/guanjq/targetdiff}.}. For each comparison model, we utilized the source code obtained from the respective repositories.

\subsection{Configuration}
All algorithms and models are developed using Python 3.8.13, with PyTorch version 1.12.1 and PyTorch Geometric version 2.5.2, under CUDA 11.0. Experiments are conducted on a server with 8 NVIDIA V100 GPUs (32 GB memory) and Intel(R) Xeon(R) Platinum 8255C CPU @ 2.50GHz. We employ a single V100 GPU for model training and leverage eight GPUs to accelerate the sampling procedure.

%% file: appendix/06_exp_results.tex
\section{More Experiment Results}
\label{more_exps}
In this section, we conduct additional experiments to further assess the effectiveness of our proposed model, particularly in addressing the \textit{separation violation issue}.

\subsection{More Results on Evaluating Separation Violation Issues}
\Cref{tab: more_results_for_collision} presents a comprehensive evaluation of \textit{violation issues} (PLVR, ALVR, MLVR) from Step-0 to Step-1000 during the inference phase. The experimental results suggest that NucleusDiff and TargetDiff have similar performance concerning violation issues in the early stages of inference~(Step-0 to Step-300). However, from Step-400 onward, NucleusDiff demonstrates a significantly faster convergence rate in addressing separation violation problems. By approximately Step-700, NucleusDiff appears to almost completely resolve the violation issues. In contrast, TargetDiff shows rapid convergence in addressing violation problems from Step-400 to Step-600, after which its performance related to violation issues shows little to no change. \Cref{tab: more_results_for_collision} offers a comprehensive overview of how these two diffusion-based models handle violation issues during the inference phase, giving us a clearer understanding of how the pretrained NucleusDiff model mitigates such problems.

\begin{figure}[ht!]
\centering
\includegraphics[width=\textwidth]{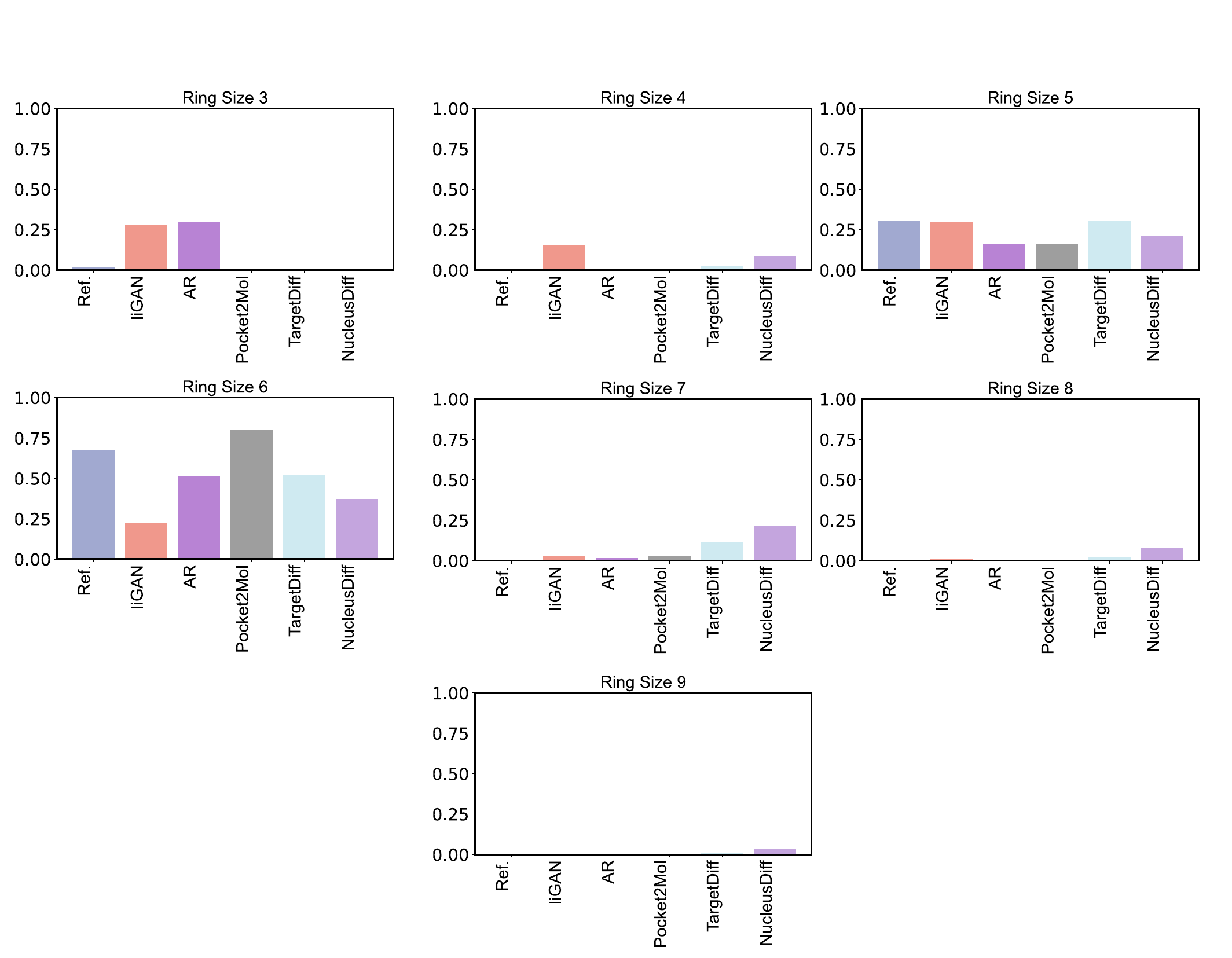}
\vspace{-2ex}
\caption{The ring size distribution of molecules generated by the baseline models and NucleusDiff.}
\label{fig:appendix:ring_size_distribution}
\end{figure}

\begin{table}[ht!]
  \centering
  \caption{The separation violation performance among pocket-ligand pairs for structure-based drug design. A lower value is better.}
  \begin{adjustbox}{max width=\textwidth}
  \begin{tabular}{l lll lll}
    \toprule[1.5pt] 
    \multirow{2}{*}{\textbf{Metrics}} & \multicolumn{3}{c}{\textbf{TargetDiff}} & \multicolumn{3}{c}{\textbf{NucleusDiff~(ours)}} \\
    \cmidrule(lr){2-4} \cmidrule(lr){5-7}
     & \textbf{PLVR} & \textbf{ALVR} & \textbf{MLVR} & \textbf{PLVR} & \textbf{ALVR} & \textbf{MLVR}  \\
    \midrule
    \textbf{Step-0}   & 17103/2300930 & 13241/230093 & 4425/10000 & 17120/2300930 & 13257/230093 & 4428/10000 \\
    \textbf{Step-100} & 11293/2300930 & 9083/230093 & 3613/10000 & 11185/2300930 & 9040/230093 & 3627/10000 \\
    \textbf{Step-200} & 6019/2300930 & 5079/230093 & 2543/10000 & 5779/2300930 & 4918/230093 & 2540/10000 \\
    \textbf{Step-300}  & 2482/2300930 & 2142/230093 & 1285/10000 & 2081/2300930 & 1847/230093 & 1254/10000 \\
    \textbf{Step-400}  & 751/2300930 & 671/230093 & 426/10000 & 444/2300930 & 414/230093 & 321/10000 \\
    \textbf{Step-500}  & 211/2300930 & 183/230093 & 124/10000 & 90/2300930 & 80/230093 & 64/10000 \\
    \textbf{Step-600}  & 84/2300930 & 77/230093 & 56/10000 & 29/2300930 & 28/230093 & 26/10000 \\
    \textbf{Step-700}  & 78/2300930 & 70/230093 & 45/10000 & 7/2300930 & 7/230093 & 7/10000 \\
    \textbf{Step-800}  & 77/2300930 & 70/230093 & 45/10000 & 4/2300930 & 4/230093 & 4/10000 \\
    \textbf{Step-900} & 78/2300930  & 70/230093& 43/10000 &  2/2300930 & 2/230093  & 2/10000\\
    \textbf{Step-1000} & 65/2300930 &60/230093 & 37/10000 & 0/2300930 & 0/230093 & 0/10000 \\
    \bottomrule[1.5pt]
  \end{tabular}
  \end{adjustbox}
  \label{tab: more_results_for_collision}
\end{table}

\subsection{The Ring Size distribution}
Many previous studies~\cite{peng2022pocket2mol, luo2022autoregressive, guan20223d} have suggested that if the ligands generated by a pretrained model exhibit high structural consistency with the ground-truth ligands in the test set, the generative model can be considered highly successful. However, we argue that this perspective is flawed. A generative model should not only learn a distribution but also possess strong generalization capabilities. Our goal is to generate a diverse array of ligands beyond the training distribution, which holds significant practical implications for drug design and discovery. In this section, we present a detailed analysis of the substructures of the molecules generated by NucleusDiff, specifically focusing on the distribution of ring sizes in the generated molecules. \Cref{fig:appendix:ring_size_distribution} illustrates the distribution of different ring sizes present in the test set, as well as in the 10,000 molecules generated by the baselines and NucleusDiff. The results reveal that, in comparison to the test set and another diffusion-based model, TargetDiff, the ring sizes in the molecules generated by these models are primarily concentrated around 5 and 6. In contrast, the ring sizes in the molecules generated by NucleusDiff are mainly distributed among 5, 6, 7, 8, and 9. Notably, the proportion of ring sizes 7, 8, and 9 is significantly higher in NucleusDiff compared to TargetDiff and the test set. This observation leads to two key insights: (1) NucleusDiff demonstrates the potential to generate more complex structures, such as intricate ring structures, compared to TargetDiff. (2) The structures generated by NucleusDiff are more novel, as evidenced by the discrepancy between the substructure distribution of the ground-truth ligands in the test set and that of the molecules generated by NucleusDiff.

\begin{table}[ht!]
\caption{The distribution of ring sizes in the test set and molecules generated by the models.}
\vspace{-10pt}
\centering
\begin{adjustbox}{max width=\textwidth}
\begin{tabular}{ccccccc}
\toprule[1.5pt]
\small{\textbf{Ring Size}} & \small{\textbf{Ref.}} & \small{\textbf{liGAN}} & \small{\textbf{AR}} & \small{\textbf{Pocket2Mol}} & \small{\textbf{TargetDiff}} & \small{\textbf{NucleusDiff~(ours)}} \\
\midrule
\textbf{3} & 1.7\%  & 28.1\% & 29.9\% & 0.1\% & 0.0\%   &0.0\%  \\
\textbf{4} & 0.0\%  & 15.7\% & 0.0\%  & 0.0\% & 2.5\%   &8.8\%  \\
\textbf{5} & 30.2\% & 29.8\% & 16.0\% & 16.4\% & 30.6\% &21.4\%  \\
\textbf{6} & 67.4\% & 22.7\% & 51.2\% & 80.4\% & 51.8\% &37.3\%  \\
\textbf{7} & 0.7\%  & 2.6\%  & 1.7\%  & 2.6\% & 11.8\%  &21.2\%  \\
\textbf{8} & 0.0\%  & 0.8\%  & 0.7\%  & 0.3\% & 2.5\%   &7.6\%  \\
\textbf{9} & 0.0\%  & 0.3\%  & 0.5\%  & 0.1\% & 0.8\%   &3.7\%  \\
\bottomrule[1.5pt]
\end{tabular}
\end{adjustbox}
\end{table}

\subsection{The Bond Distribution}
In our study, we evaluate the performance of various generative models in terms of their ability to reproduce the bond distributions observed in reference molecules. Our primary focus was on the NucleusDiff model, which demonstrated a unique capability in generating diverse molecular substructures.

\begin{figure}[ht!]
\centering
\includegraphics[width=\textwidth]{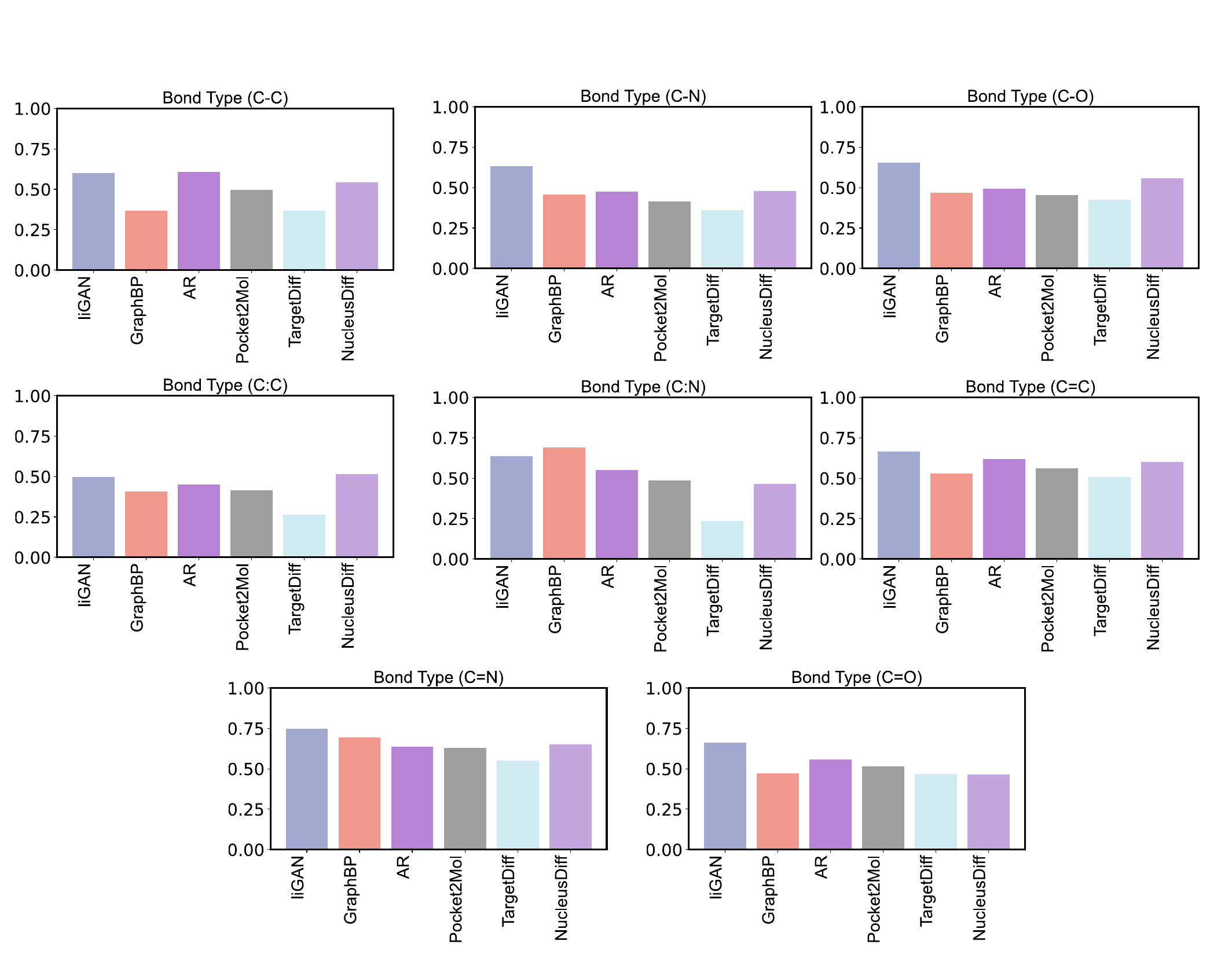}
\vspace{-2ex}
\caption{The bond distribution of the molecules generated by baseline models and NucleusDiff.}
\label{fig:appendix:bond_distribution}
\end{figure}

\Cref{fig:appendix:bond_distribution} presents a comparative analysis of the bond distributions for different models, including liGAN, GraphBP, AR, Pocket2Mol, TargetDiff, and NucleusDiff. The NucleusDiff model consistently shows a balanced distribution across various bond types (C-C, C=C, C-N, C=N, C-O, C=O, C:C, C:N), indicating its proficiency in capturing the structural diversity inherent in the reference dataset.

\Cref{tab:bond distribution} provides quantitative insights through the Jensen-Shannon divergence between the bond distance distributions of reference molecules and those generated by each model. The NucleusDiff model exhibits competitive divergence values across all bond types, highlighting its effectiveness in mimicking the bond length distributions of real molecules. The Jensen-Shannon divergence is a measure of similarity between two probability distributions, with values ranging from 0 to 1. Lower values indicate higher similarity between distributions. In this context, the low divergence values achieved by NucleusDiff, particularly for bonds like C=N (0.649) and C=O (0.464), signify that the bond length distributions in the generated molecules closely resemble those in the reference molecules. This high degree of similarity suggests that NucleusDiff can accurately capture and reproduce the structural characteristics of real molecules. Notably, for bonds like C=N and C=O, NucleusDiff achieves divergence values of 0.649 and 0.464, respectively, which are relatively low and suggest a high degree of similarity to the reference distributions.

The superior performance of NucleusDiff in generating diverse substructures can be attributed to its advanced architectural design, which allows for fine-grained control over molecular features. While other models like liGAN and GraphBP show competence in certain aspects of molecular generation, they often struggle with maintaining consistent performance across various bond types. For instance, liGAN exhibits higher divergence values for most bond types, indicating less accurate reproduction of bond distributions. GraphBP, while performing well for some bonds, shows inconsistency across different bond types. In contrast, NucleusDiff demonstrates a more balanced and consistently low divergence across all bond types, highlighting its superior capability in generating diverse and structurally accurate molecules. This is evident from its ability to generate molecules with a wide range of bond types, maintaining a high degree of structural fidelity to natural molecules. Consequently, NucleusDiff not only ensures the generation of chemically valid molecules but also enhances the exploration of the chemical space by producing a variety of substructures, which is crucial for applications in drug discovery and materials science.

In summary, the NucleusDiff model stands out in its ability to generate molecular substructures with considerable diversity, closely mirroring the bond distributions of reference molecules. This capability underscores its potential as a powerful tool for the generation of novel and diverse molecular entities.

\begin{table}[ht!]
\caption{The Jensen-Shannon divergence between the distributions of bond distances for reference versus generated molecules is analyzed. In this context, "-", "=", and ":" denote single, double, and aromatic bonds, respectively.}
\vspace{-10pt}
\centering
\begin{adjustbox}{max width=\textwidth}
\begin{tabular}{ccccccc} 
\toprule[1.5pt]
\small{\textbf{Bond}} & \small{\textbf{liGAN}} & \small{\textbf{GraphBP}} & \small{\textbf{AR}} & \small{\textbf{Pocket2Mol}} & \small{\textbf{TargetDiff}} & \small{\textbf{NucleusDiff~(ours)}} \\
\midrule
\textbf{C$-$C} & 0.601 & 0.368 & 0.609 & 0.496 & 0.367    &0.544  \\
\textbf{C$=$C} & 0.665 & 0.530 & 0.620 & 0.561 & 0.507    &0.599  \\
\textbf{C$-$N} & 0.634 & 0.456 & 0.474 & 0.416 & 0.361    &0.478 \\
\textbf{C$=$N} & 0.749 & 0.693 & 0.635 & 0.629 & 0.551    &0.649  \\
\textbf{C$-$O} & 0.656 & 0.467 & 0.492 & 0.454 & 0.424    &0.559 \\
\textbf{C$=$O} & 0.661 & 0.471 & 0.558 & 0.516 & 0.467    &0.464  \\
\textbf{C$:$C} & 0.497 & 0.407 & 0.451 & 0.416 & 0.264    &0.517 \\
\textbf{C$:$N} & 0.638 & 0.689 & 0.552 & 0.487 & 0.234    &0.464  \\
\bottomrule[1.5pt]
\end{tabular}
\end{adjustbox}
\label{tab:bond distribution}
\end{table}

\begin{figure}[ht!]
\centering
\includegraphics[width=\textwidth]{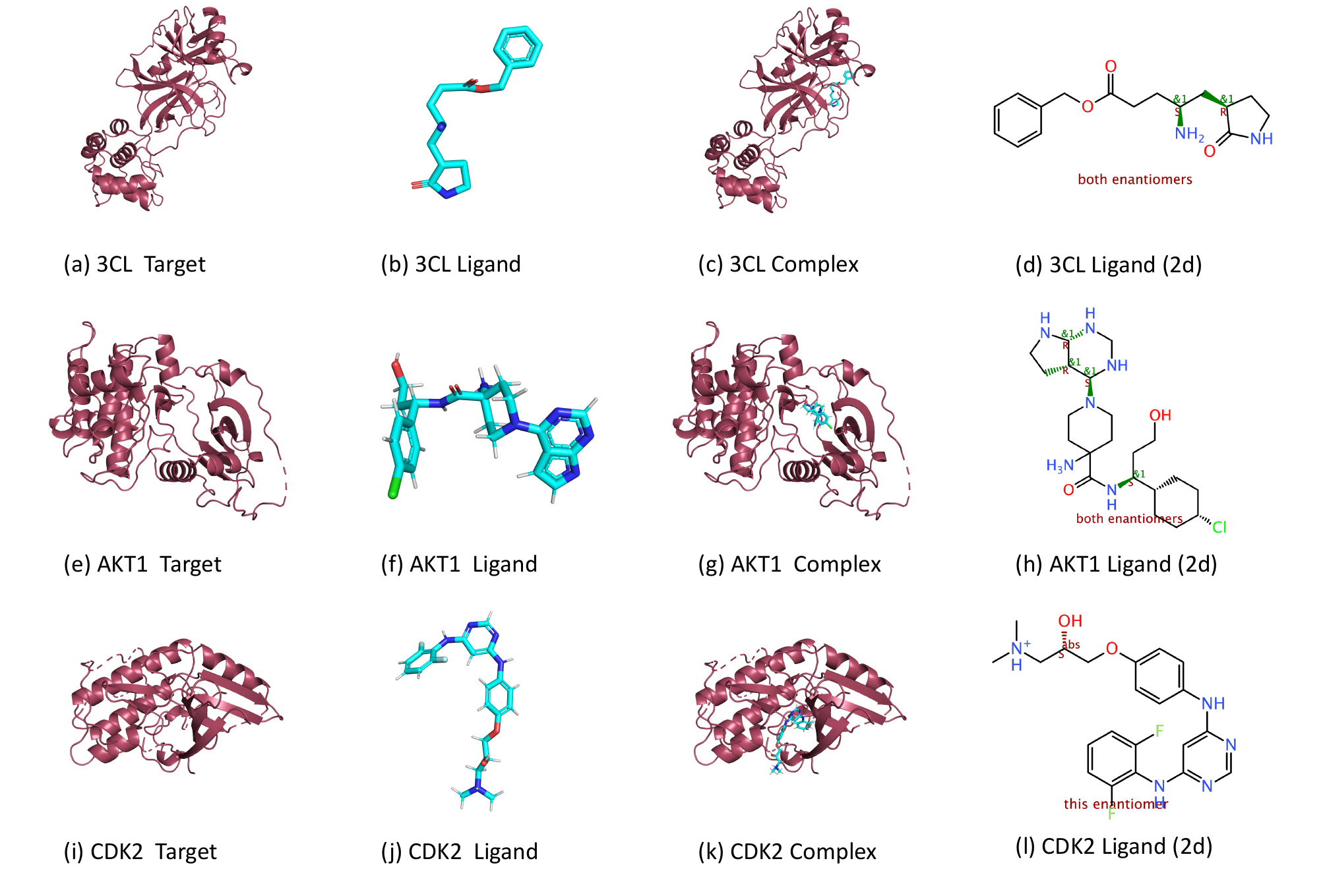}
\vspace{-2ex}
\caption{Visualization for drug design on COVID-19 and the other two therapeutic targets.}
\label{fig:drug_design_on_covid_19}
\end{figure}

\subsection{More Experiments for Advances in Drug Design for COVID-19 and Other Therapeutic Targets}
\label{appendix: Advances in Drug Design for COVID-19 and Other Therapeutic Targets}

While evaluating a model's performance on representative test sets is a common practice in the machine learning community, scientists and drug designers are more interested in the models' performance in real-world applications. Therefore, following prior work~\cite{zhang2023learning}, we conduct more experiments on a carefully curated dataset comprising over three out-of-distribution therapeutic targets, including those related to COVID-19, along with their experimentally validated active ligands. This dataset is designed to more accurately reflect the models' potential in practical settings. In this context, we generate molecules targeting these specific therapeutic targets and subsequently evaluate their binding affinity, drug-likeness properties, and similarity to known active compounds. Additionally, we assess the performance of NucleusDiff and TargetDiff on the three therapeutic targets (including the COVID-19 target) using the proposed violation metrics. This step is critical for comprehensively analyzing whether NucleusDiff can effectively address violation issues in structure-based drug design for real-world therapeutic targets, which is a central focus of our research.

\paragraph{Experimental Setup.} 
Here, we employ the pretrained models of TargetDiff and \model{} to sample 1000 ligands for each target: \textbf{3CL}, \textbf{ATK1}, and \textbf{CDK2}. Both models utilize diffusion-based methodologies, facilitating the generation of molecular structures through iterative refinements. For each target, the sampling process is conducted over 1000 timesteps to ensure the convergence of molecular structures. The generated ligands are evaluated using several metrics to comprehensively assess their performance. Specifically, we use AutoDock Vina to compute binding affinity metrics, including the Vina Score, Vina Min, and Vina Dock. The Vina Score measures the overall binding affinity, while Vina Min and Vina Dock provide insights into the minimum and docking scores, respectively, indicating the strength of ligand-target interactions. High Affinity metrics are also calculated to gauge the proportion of generated ligands with exceptionally strong binding affinity. Drug-likeness properties are assessed using the Quantitative Estimate of Drug-likeness (QED) score, which evaluates the likelihood of a compound to exhibit drug-like characteristics. Additionally,  we measure synthetic accessibility (SA), which quantifies the ease with which a compound can be synthesized, and molecular diversity, which assesses the range and heterogeneity of structures within the generated ligand set. To further analyze the models' performance in addressing separation violations, we examine separation violation ratios at 11 timesteps, sampled every 100 timesteps from Step-0 to Step-1000. The metrics used for this analysis included PLVR, ALVR, and MLVR. These metrics provide insights into how the separation violation issue evolved during the inference process, with lower values indicating fewer violations and better structural compatibility. This comprehensive experimental setup allows for a rigorous evaluation of NucleusDiff and TargetDiff, highlighting their performance in generating high-quality ligands for the specified therapeutic targets. The results, detailed in the subsequent sections and tables, underscore the strengths of each model and identify potential areas for improvement in the context of structure-based drug design.

\begin{figure}[ht!]
\centering
\includegraphics[width=\textwidth]{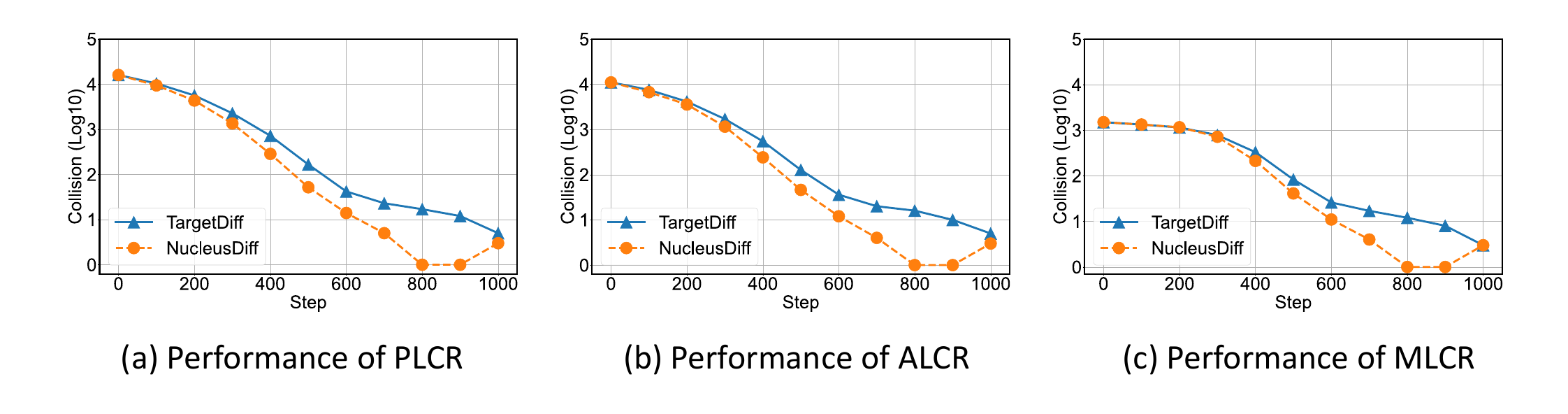}
\vspace{-2ex}
\caption{Visualization of the separation violation ratios for TargetDiff and NucleusDiff across three targets (3CL, AKT1, CDK2).}
\label{fig2:vis_of_out_of_distribution_metric_performance}
\end{figure}

\begin{table}[t!]
  \caption{The separation violation performance among pocket-ligand pairs for structure-based drug design in the three real-world therapeutic targets. A lower value is better.}
  \vspace{-10pt}
    \centering
  \begin{adjustbox}{max width=\textwidth}
  \begin{tabular}{l lll lll}
    \toprule[1.5pt]
     \multirow{2}{*}{\textbf{Metrics}} & \multicolumn{3}{c}{\textbf{TargetDiff}} & \multicolumn{3}{c}{\textbf{NucleusDiff~(ours)}} \\
    \cmidrule(lr){2-4} \cmidrule(lr){5-7}
     & \textbf{PLVR} & \textbf{ALVR} & \textbf{MLVR} & \textbf{PLVR} & \textbf{ALVR} & \textbf{MLVR}  \\
    \midrule
    \textbf{Step-0}    &15931/810000   &10979/81000   &1496/3000   &15924/810000   &10973/81000   &1499/3000   \\
    \textbf{Step-100}  &10327/810000   &7490/81000    &1329/3000   &9369/810000    &6673/81000    &1328/3000   \\
    \textbf{Step-200}  &5611/810000    &4125/81000    &1147/3000   &4329/810000    &3556/81000    &1158/3000   \\
    \textbf{Step-300}  &2258/810000    &1695/81000    &779/3000    &1345/810000    &1151/81000    &716/3000    \\
    \textbf{Step-400}  &719/810000     &547/81000     &330/3000    &285/810000     &241/81000     &212/3000    \\
    \textbf{Step-500}  &166/810000     &127/81000     &83/3000     &52/810000      &46/81000      &41/3000     \\
    \textbf{Step-600}  &42/810000      &36/81000      &26/3000     &14/810000      &12/81000      &11/3000     \\
    \textbf{Step-700}  &23/810000      &20/81000      &17/3000     &5/810000       &4/81000       &4/3000      \\
    \textbf{Step-800}  &17/810000      &16/81000      &12/3000     &1/810000       &1/81000       &1/3000      \\
    \textbf{Step-900}  &12/810000      &10/81000      &8/3000      &1/810000       &1/81000       &1/3000      \\
    \textbf{Step-1000} &5/810000       &5/81000       &3/3000      &3/810000       &3/81000       &3/3000      \\
    \bottomrule[1.5pt]
  \end{tabular}
  \end{adjustbox}
  \label{tab2: combined_results_for_collision}
\end{table}

\begin{table}[t!]
\centering
\caption{A summary of 14 biochemical properties for molecules generated by TargetDiff and \model{} for \textbf{target 3CL}. The symbols ($\uparrow$) and ($\downarrow$) indicate whether a higher or lower value is preferable for each property. 
}
\vspace{-2ex}
\begin{adjustbox}{max width=\textwidth}
\begin{tabular}{c cc cc cc cc cc cc cc}
\toprule[1.5pt]
\multirow{2}{*}{\textbf{Metrics}} & \multicolumn{2}{c}{\textbf{Vina Score ($\downarrow$)}} & \multicolumn{2}{c}{\textbf{Vina Min ($\downarrow$)}} & \multicolumn{2}{c}{\textbf{Vina Dock ($\downarrow$)}} & \multicolumn{2}{c}{\textbf{High Affinity ($\uparrow$)}} & \multicolumn{2}{c}{\textbf{QED ($\uparrow$)}}   & \multicolumn{2}{c}{\textbf{SA ($\uparrow$)}} & \multicolumn{2}{c}{\textbf{Diversity ($\uparrow$)}} \\
\cmidrule(lr){2-3} \cmidrule(lr){4-5} \cmidrule(lr){6-7} \cmidrule(lr){8-9} \cmidrule(lr){10-11} \cmidrule(lr){12-13} \cmidrule(lr){14-15} 
 & \textbf{Avg.} & \textbf{Med.} & \textbf{Avg.} & \textbf{Med.} & \textbf{Avg.} & \textbf{Med.} & \textbf{Avg.} & \textbf{Med.} & \textbf{Avg.} & \textbf{Med.} & \textbf{Avg.} & \textbf{Med.} & \textbf{Avg.} & \textbf{Med.}\\
\midrule
\textbf{TargetDiff} & -4.82   &-5.08  & -5.61  & -5.68 &-6.39  & -6.49 & 50.5\%  &50.5\% & \textbf{0.56}     & \textbf{0.54} &\textbf{0.62}  &   \textbf{0.61}  & \textbf{0.76}    & \textbf{0.76}      \\
\midrule
\textbf{NucleusDiff~(ours)} &\textbf{-5.85}  &\textbf{-5.80}  &\textbf{-6.21}  &\textbf{-6.23}  & \textbf{-6.74} & \textbf{-6.84}  & \textbf{70.0\%} & \textbf{70.0\%}    & 0.43     &0.42  & 0.54 &0.53   & 0.73&0.73  \\
\bottomrule[1.5pt]
\end{tabular}
\end{adjustbox}
\label{tab:mol_prop_3cl}
\end{table}

\begin{table}[t!]
\centering
\caption{A summary of 14 biochemical properties for molecules generated by TargetDiff and \model{} for \textbf{target AKT1}. The symbols ($\uparrow$) and ($\downarrow$) indicate whether a higher or lower value is preferable for each property. 
}
\vspace{-2ex}
\begin{adjustbox}{max width=\textwidth}
\begin{tabular}{ccc cc cc cc cc cc cc}
\toprule[1.5pt]
\multirow{2}{*}{\textbf{Metrics}} & \multicolumn{2}{c}{\textbf{Vina Score ($\downarrow$)}} & \multicolumn{2}{c}{\textbf{Vina Min ($\downarrow$)}} & \multicolumn{2}{c}{\textbf{Vina Dock ($\downarrow$)}} & \multicolumn{2}{c}{\textbf{High Affinity ($\uparrow$)}} & \multicolumn{2}{c}{\textbf{QED ($\uparrow$)}}   & \multicolumn{2}{c}{\textbf{SA ($\uparrow$)}} & \multicolumn{2}{c}{\textbf{Diversity ($\uparrow$)}} \\
\cmidrule(lr){2-3} \cmidrule(lr){4-5} \cmidrule(lr){6-7} \cmidrule(lr){8-9} \cmidrule(lr){10-11} \cmidrule(lr){12-13} \cmidrule(lr){14-15} 
 & \textbf{Avg.} & \textbf{Med.} & \textbf{Avg.} & \textbf{Med.} & \textbf{Avg.} & \textbf{Med.} & \textbf{Avg.} & \textbf{Med.} & \textbf{Avg.} & \textbf{Med.} & \textbf{Avg.} & \textbf{Med.} & \textbf{Avg.} & \textbf{Med.}\\
\midrule
\textbf{TargetDiff}  & -8.31  & -8.22   & -8.77 &-8.69  & -9.24 &-9.17 & 34.2\% &34.2\%  & \textbf{0.43}    & \textbf{0.42} &\textbf{0.51}  &   \textbf{0.52}  & 0.55    & 0.55      \\
\midrule
\textbf{NucleusDiff~(ours)} &\textbf{-9.42}  &\textbf{-9.37}  &\textbf{-9.33}  &\textbf{-9.39}  & \textbf{-9.84} & \textbf{-9.80}  & \textbf{52.8\%} & \textbf{52.8\%}    & 0.29     &0.27 & 0.41 &0.43   & \textbf{0.56}&\textbf{0.56}  \\
\bottomrule[1.5pt]
\end{tabular}
\end{adjustbox}
\label{tab:mol_prop_akt1}
\end{table}

\begin{table}[t!]
\centering
\caption{A summary of 14 biochemical properties for molecules generated by TargetDiff and \model{} for \textbf{target CDK2}. The symbols ($\uparrow$) and ($\downarrow$) indicate whether a higher or lower value is preferable for each property.
}
\vspace{-2ex}
\begin{adjustbox}{max width=\textwidth}
\begin{tabular}{c cc cc cc cc cc cc cc}
\toprule[1.5pt]
\multirow{2}{*}{\textbf{Metrics}} & \multicolumn{2}{c}{\textbf{Vina Score ($\downarrow$)}} & \multicolumn{2}{c}{\textbf{Vina Min ($\downarrow$)}} & \multicolumn{2}{c}{\textbf{Vina Dock ($\downarrow$)}} & \multicolumn{2}{c}{\textbf{High Affinity ($\uparrow$)}} & \multicolumn{2}{c}{\textbf{QED ($\uparrow$)}}   & \multicolumn{2}{c}{\textbf{SA ($\uparrow$)}} & \multicolumn{2}{c}{\textbf{Diversity ($\uparrow$)}} \\
\cmidrule(lr){2-3} \cmidrule(lr){4-5} \cmidrule(lr){6-7} \cmidrule(lr){8-9} \cmidrule(lr){10-11} \cmidrule(lr){12-13} \cmidrule(lr){14-15} 
 & \textbf{Avg.} & \textbf{Med.} & \textbf{Avg.} & \textbf{Med.} & \textbf{Avg.} & \textbf{Med.} & \textbf{Avg.} & \textbf{Med.} & \textbf{Avg.} & \textbf{Med.} & \textbf{Avg.} & \textbf{Med.} & \textbf{Avg.} & \textbf{Med.}\\
\midrule
\textbf{TargetDiff}& -8.96   &-8.93 & -9.43   & -9.39  &-9.91 & -9.87 & \textbf{97.0\%}  & \textbf{97.0\%} & \textbf{0.52}     & \textbf{0.52} & \textbf{0.50}  &  \textbf{0.51}  & 0.55    & 0.55      \\
\midrule
\textbf{NucleusDiff~(ours)} &\textbf{-10.78}  &\textbf{-11.02}  &\textbf{-10.71}  &\textbf{-11.10}  & \textbf{-11.10} & \textbf{-11.30}  & 95.7\% & 95.7\%    & 0.38     &0.38  & 0.38 &0.36   & \textbf{0.59}&\textbf{0.59}  \\
\bottomrule[1.5pt]
\end{tabular}
\end{adjustbox}
\label{tab:mol_prop_cdk2}
\end{table}

\paragraph{Separation Violation Evaluation.} 
To evaluate the performance of NucleusDiff, we conduct a comprehensive comparison with TargetDiff, focusing on their application to real-world scenarios, including therapeutic targets such as COVID-19. Both methods leverage diffusion-based models, which enable a thorough understanding of the separation violation issue during the inference process of DDPM~\cite{ho2020denoising}. Our datasets include three real-world therapeutic targets (3CL, AKT1, CDK2) with experimentally validated active ligands. We analyze three key metrics for separation violation at 11 timesteps, sampled every 100 timtsteps from Step-0 to Step-1000. The main results are summarized in \Cref{tab2: combined_results_for_collision}, which presents the violation metrics for each method across these steps. In \Cref{tab2: combined_results_for_collision}, we observe that TargetDiff shows a significant reduction in atomic violations from Step-0 to Step-1000. Conversely, NucleusDiff maintains a consistently lower violation ratio over the same inference steps. Notably, NucleusDiff significantly outperforms TargetDiff across all three violation metrics (PLCR, ALCR, and MLCR), with the difference approaching an order of magnitude. Specifically, in the final sampling steps, NucleusDiff achieves an almost negligible violation ratio, underscoring its superior performance. Referring to~\Cref{fig2:vis_of_out_of_distribution_metric_performance}, which visualizes the separation violation ratio for both methods across three targets (3CL, AKT1, CDK2), we can see a marked contrast in the convergence trends. NucleusDiff demonstrates a more pronounced and rapid reduction in violation ratios compared to TargetDiff. This suggests that NucleusDiff is better suited to real-world applications, especially for real-world therapeutic targets. The enhanced convergence and lower violation ratios of NucleusDiff highlight its potential for practical deployment in drug design.

\paragraph{Binding Affinity Evaluation.}  For each target protein, we generate 1000 ligand molecules, resulting in a total of 3000 molecules per model. The comprehensive results for NucleusDiff and TargetDiff are displayed in \Cref{tab:mol_prop_3cl}, \Cref{tab:mol_prop_akt1}, and \Cref{tab:mol_prop_cdk2}. We note that NucleusDiff outperforms TargetDiff in 8 out of the 14 evaluated metrics across all three therapeutic targets. According to the Vina Score, NucleusDiff achieves an average score of -5.85 for 3CL, -9.42 for AKT1, and -10.78 for CDK2, indicating superior binding affinity compared to TargetDiff. Similarly, for the Vina Min and Vina Dock metrics, NucleusDiff consistently outperforms TargetDiff, suggesting that NucleusDiff is more effective in predicting highly favorable binding poses. In terms of High Affinity, NucleusDiff exhibits a significant advantage over TargetDiff, with 70.0\% for 3CL, 52.8\% for AKT1, and 95.7\% for CDK2, compared to 50.5\%, 34.2\%, and 97.0\% respectively for TargetDiff. This indicates that NucleusDiff is more proficient at generating ligands with strong binding interactions. Regarding the QED metric, although NucleusDiff's scores are slightly lower than those of TargetDiff, it still maintains acceptable drug-likeness properties. For instance, the QED scores for NucleusDiff are 0.43 for 3CL, 0.29 for AKT1, and 0.38 for CDK2, compared to TargetDiff's 0.56, 0.43, and 0.52 respectively. The SA scores for NucleusDiff are within a reasonable range, ensuring that the generated molecules are synthetically accessible. The Diversity metric further underscores the capability of NucleusDiff to explore a broader chemical space, with scores of 0.73, 0.56, and 0.59 for 3CL, AKT1, and CDK2 targets respectively, compared to TargetDiff's 0.76, 0.55, and 0.55. These results collectively demonstrate that NucleusDiff not only excels in generating high-affinity ligands but also maintains a balance between drug-likeness, synthetic accessibility, and structural diversity. Consequently, NucleusDiff shows great potential for generating viable drug candidates in real-world therapeutic settings, including those for COVID-19.

\begin{figure}[tb]
\centering
\includegraphics[width=\textwidth]{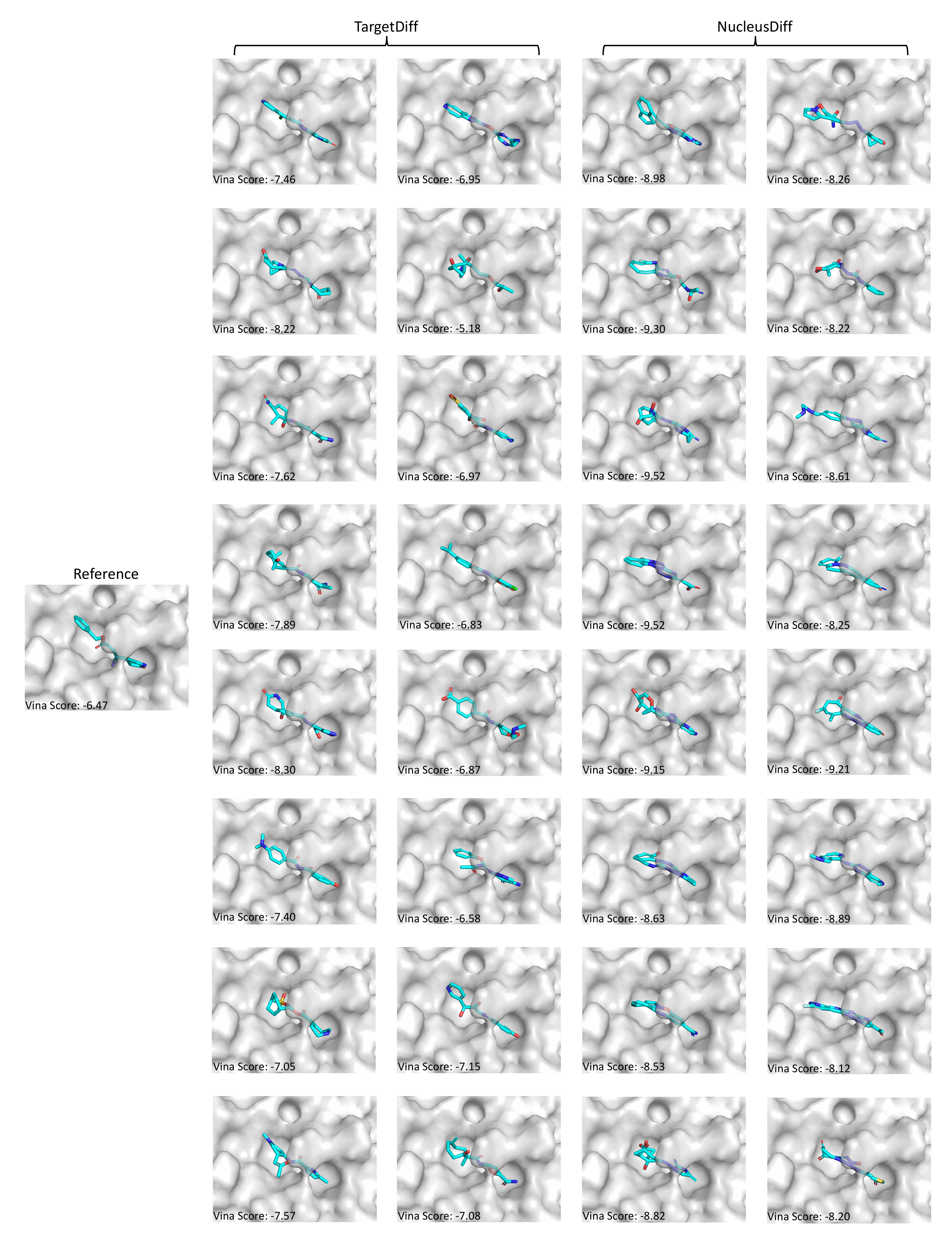}
\vspace{-2ex}
\caption{Visualization of the generated molecules by TargetDiff and NucleusDiff for \textbf{target 3CL}.}
\label{fig:vis_of_3cl}
\end{figure}

\begin{figure}[tb]
\centering
\includegraphics[width=\textwidth]{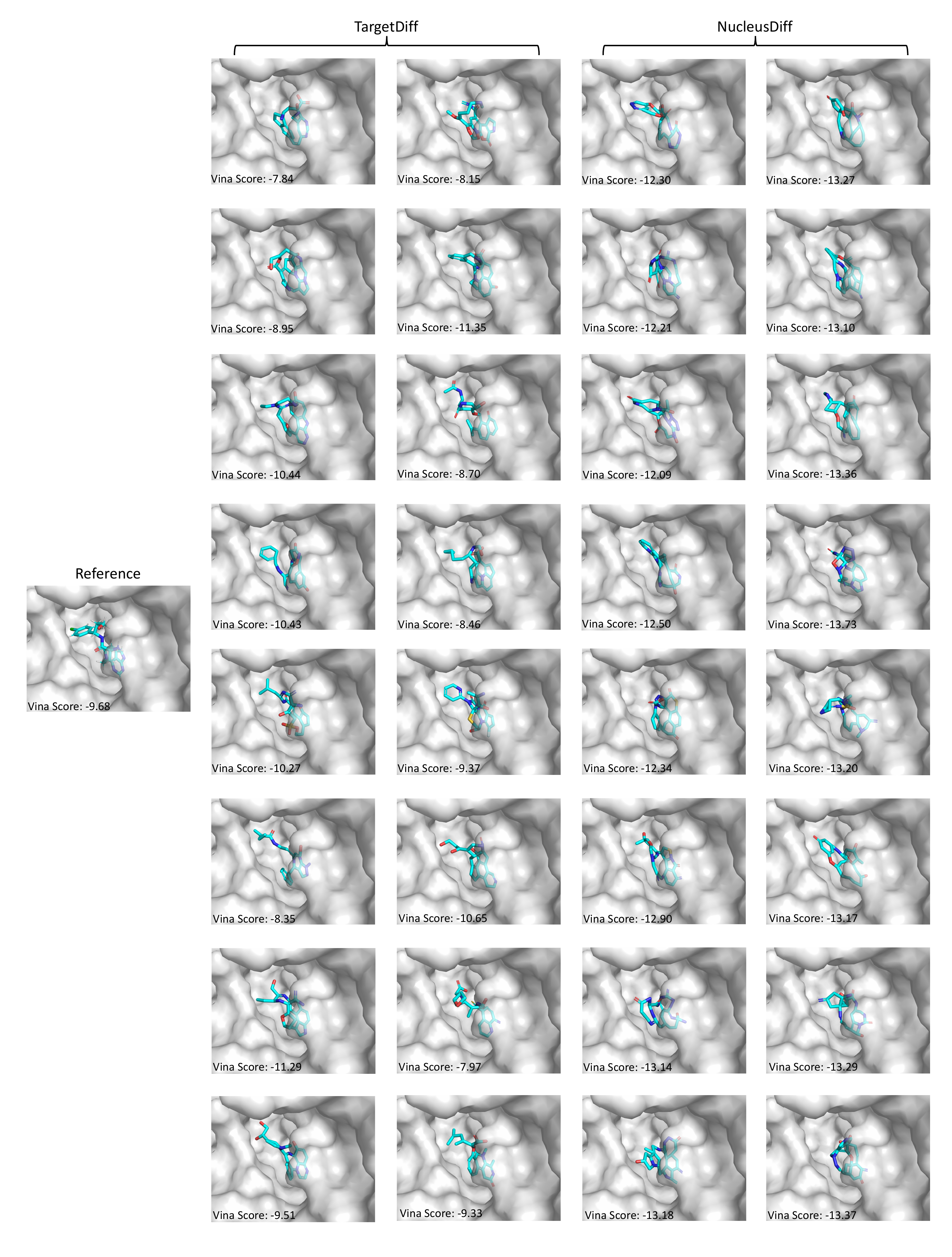}
\vspace{-2ex}
\caption{Visualization of the generated molecules by TargetDiff and NucleusDiff for \textbf{target AKT1}.}
\label{fig:vis_of_akt1}
\end{figure}

\begin{figure}[tb]
\centering
\includegraphics[width=\textwidth]{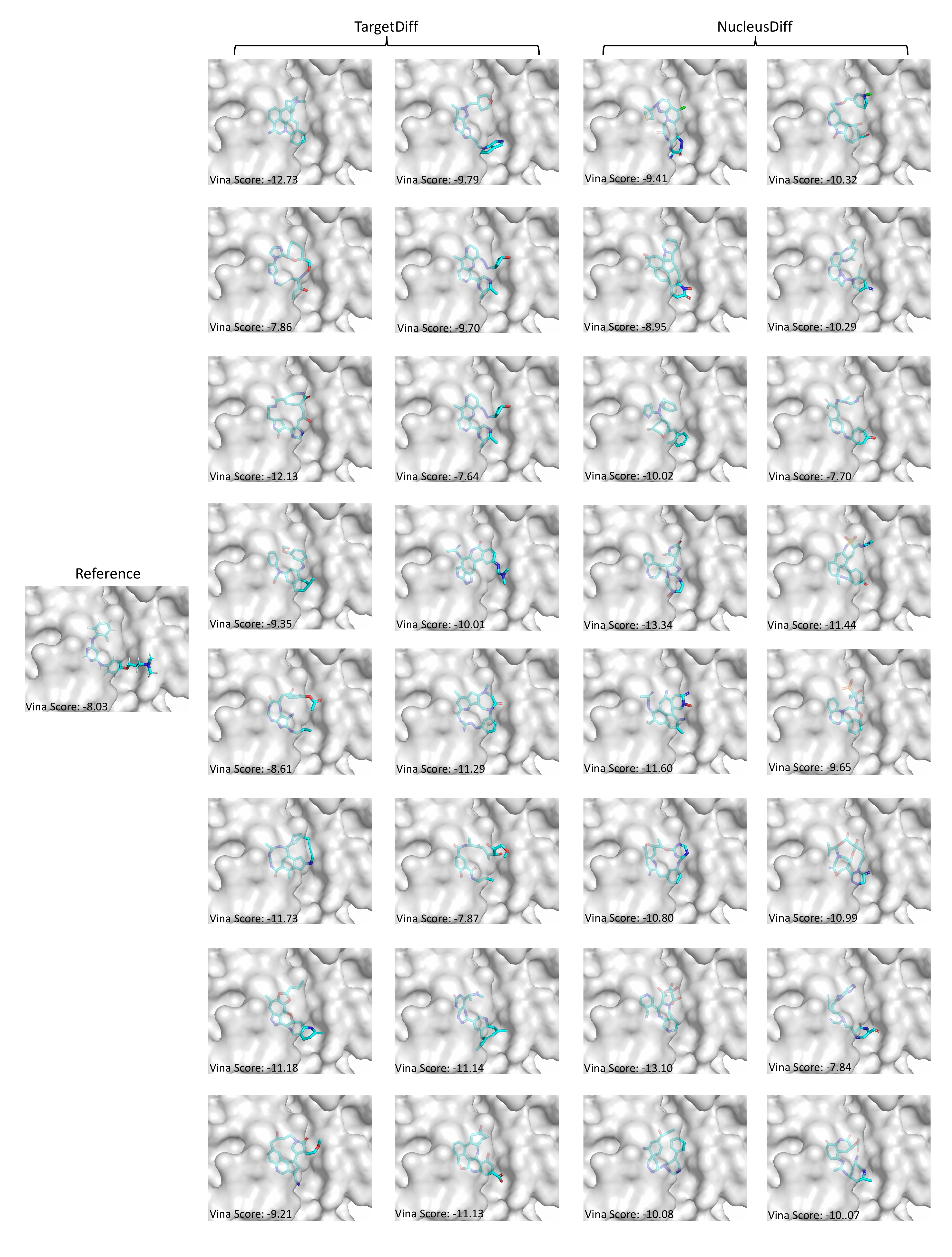}
\vspace{-2ex}
\caption{Visualization of the generated molecules by TargetDiff and NucleusDiff for \textbf{target CDK2}.}
\label{fig:vis_of_cdk2}
\end{figure}

\paragraph{Visual Analysis of NucleusDiff.} \Cref{fig:vis_of_3cl}, \Cref{fig:vis_of_akt1}, and \Cref{fig:vis_of_cdk2} illustrate the visual representation of the ligands generated by NucleusDiff and TargetDiff for several therapeutic targets, including COVID-19 targets such as 3CL, AKT1, and CDK2. We select these targets due to their significance in current drug design efforts and to evaluate the robustness of the models in generating viable drug candidates. We observe that both TargetDiff and NucleusDiff demonstrate the potential to generate stable and biologically relevant structures. From the perspective of binding affinity, as indicated by the Vina Scores, NucleusDiff tends to produce ligands with higher affinity compared to those generated by TargetDiff. This trend is specifically evident in the ligands generated for the targets 3CL, AKT1, and CDK2, as shown in \Cref{fig:vis_of_3cl}, \Cref{fig:vis_of_akt1}, and \Cref{fig:vis_of_cdk2}. When considering separation violations in the generated molecules, a more detailed comparison between TargetDiff and NucleusDiff can be made. For instance, with the 3CL target, it is evident that the ligands generated by TargetDiff exhibit a less defined positioning relative to the binding pocket, leading to potential separation violations and suboptimal binding interactions. In contrast, the ligands generated by NucleusDiff show a much clearer boundary and more precise alignment with the protein pockets. This indicates that NucleusDiff learns both the relative positioning of ligands in protein pockets and the physical rules governing atomic and electronic distributions within ligands. Moreover, NucleusDiff's ability to generate ligands with higher binding affinities and better structural compatibility suggests its superior performance in real-world drug design scenarios. This is particularly crucial for rapidly evolving therapeutic targets such as those related to COVID-19, where the accuracy and efficiency of ligand generation can significantly impact the speed of drug development. In conclusion, the visual and quantitative analyses confirm that NucleusDiff outperforms TargetDiff in generating ligands with higher affinity and better structural compatibility for critical therapeutic targets. This enhanced performance underscores the potential of NucleusDiff in facilitating more effective and rapid drug discovery processes.

\clearpage
\subsection{More Visualization Results}  

\begin{figure}[ht!]
\centering
\includegraphics[width=\textwidth]{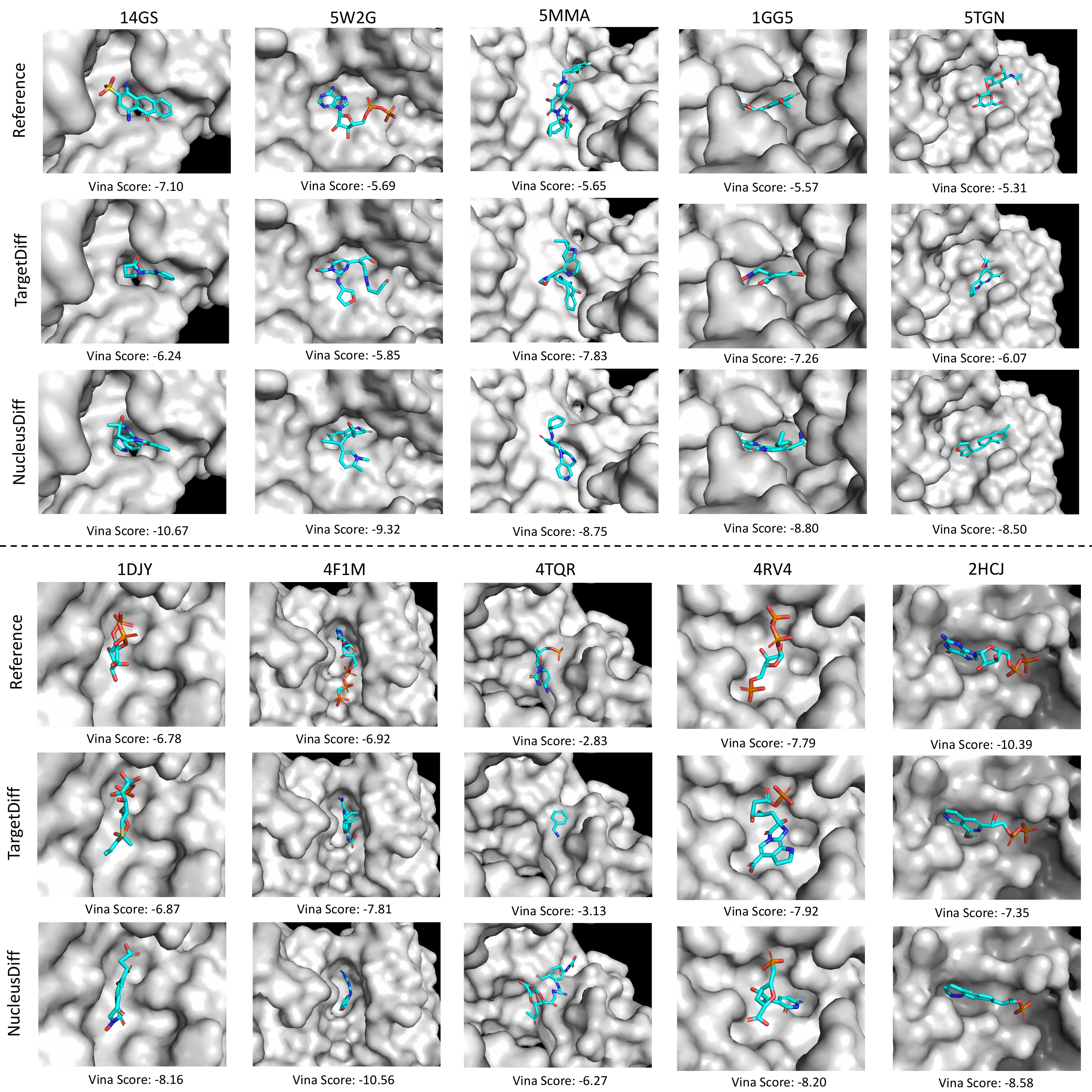}
\vspace{-2ex}
\caption{More visualization of the generated molecules by TargetDiff and NucleusDiff.}
\label{fig:more visualization}
\end{figure}

Here, we visualize additional molecules generated by NucleusDiff and employ TargetDiff as a comparative framework to illustrate the spatial configurations of the synthesized molecules. \Cref{fig:more visualization} presents additional visualizations of ligands generated by TargetDiff and \model{}, compared against the reference ligands for several protein pockets. The selected protein pockets for this analysis include 14GS, 5W2G, 2RHY, 1GGS, 5TGN, 1DJY, 4F1M, 4TQR, 5NGZ, and 1K9T. The Vina scores demonstrate that NucleusDiff consistently generates ligands with higher binding affinities compared to both TargetDiff and the reference ligands across all analyzed protein pockets. We can group our findings as follows:

\begin{itemize}
    \item For protein pockets 14GS, 5W2G, 2RHY, and 1GGS, NucleusDiff achieves significantly lower Vina scores, indicating stronger binding affinities. Specifically, NucleusDiff's scores for these pockets are -10.67, -9.32, -7.59, and -8.80 respectively, compared to TargetDiff's scores of -6.24, -5.85, -3.67, and -7.26.
    
    \item Similarly, for protein pockets 5TGN, 1DJY, 4F1M, 4TQR, 5NGZ, and 1K9T, NucleusDiff demonstrates superior performance in terms of binding affinity, consistently achieving lower Vina scores compared to TargetDiff and the reference ligands.
\end{itemize}

In addition to improved Vina scores, NucleusDiff also exhibits superior accuracy in the positional relationship between the ligands and the protein pockets, reducing the risk of separation violation. This advantage is particularly evident in the following cases:

\begin{itemize}
    \item For protein pockets 1DJY, 4F1M, and 4TQR, the ligands generated by NucleusDiff are more accurately positioned within the pockets, with clearer boundaries and better adherence to physical atomic interaction rules.
    
    \item In contrast, TargetDiff's generated ligands often exhibit less clear positioning and potential separation violations across various protein pockets.
\end{itemize}

The enhanced performance of NucleusDiff in both binding affinity and positional accuracy underscores its potential advantages in drug design and the modeling of protein-ligand interactions. These results highlight NucleusDiff's ability to generate stable and precisely positioned ligands, making it a more effective tool for computational drug discovery and optimization processes.

\subsection{Visualization of the Training Process}  

\begin{figure}[ht!]
\centering
\includegraphics[width=\textwidth]{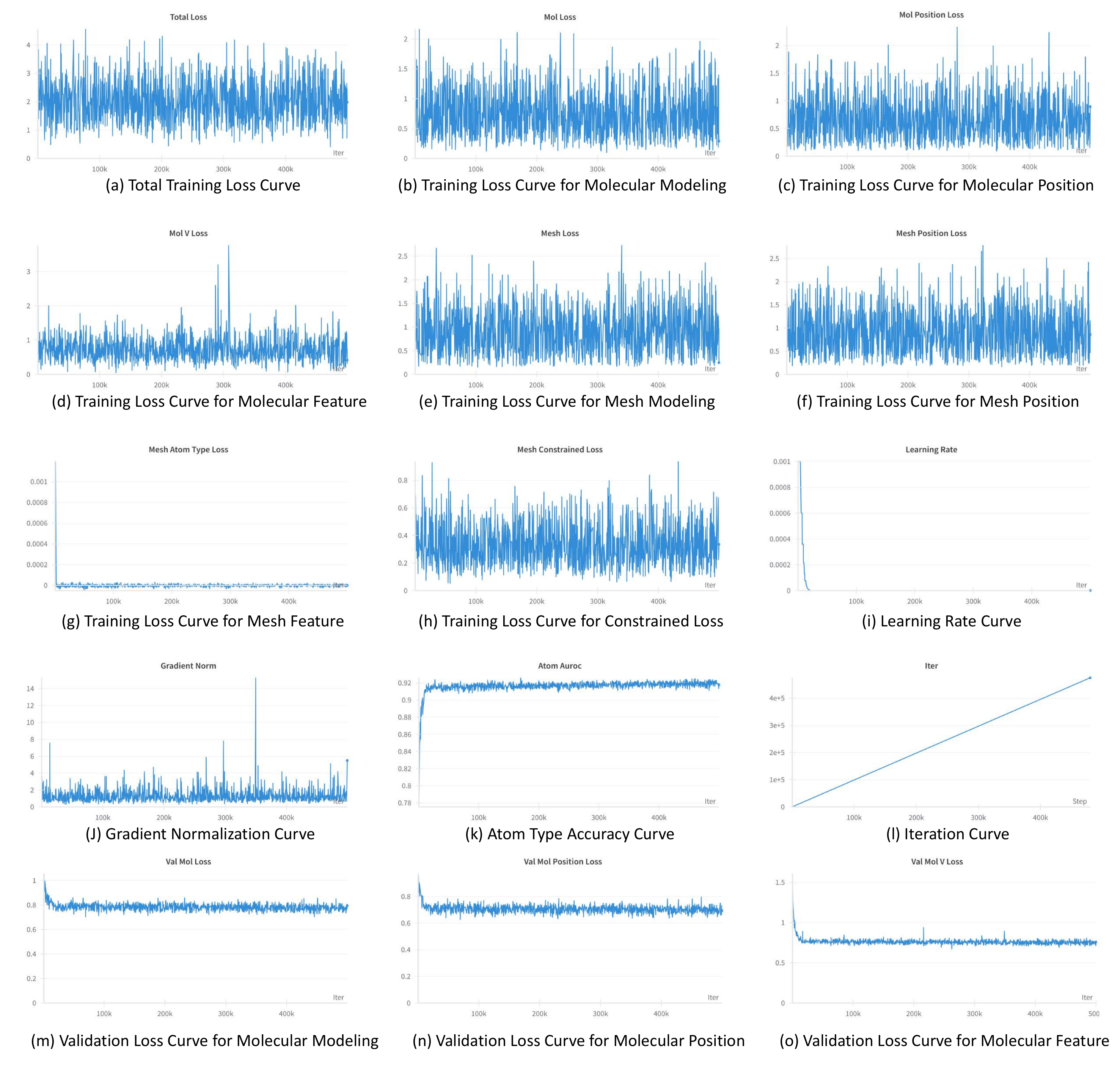}
\vspace{-2ex}
\caption{The visualization of the training process of \model{}.}
\label{fig: visualization of the training process}
\end{figure}

The visualizations in \Cref{fig: visualization of the training process} provide insights into the training dynamics of the model. The Total Training Loss Curve (a) shows overall loss with noticeable fluctuations but a general downward trend, indicating that the model learns over time. The training loss curves for molecular and mesh modeling (b-f) exhibit significant variability, suggesting challenges in these specific tasks. In contrast, the Training Loss Curve for Mesh Feature (g) displays lower values and less fluctuation, indicating that the model finds it easier to learn mesh features. The Constrained Loss curve (h) shows fewer oscillations, implying that constraints help stabilize the training process. The Learning Rate Curve (i) remains relatively constant, suggesting a stable learning rate policy. The Gradient Normalization Curve (j) with occasional spikes indicates moments of large gradient changes, while the Atom Type Accuracy Curve (k) shows stable accuracy, reflecting consistent performance in predicting atom types. The Iteration Curve (l) linearly increases, reflecting the progression of training steps over time. The validation loss curves (m-o) mirror the training loss curves, showing high variability but general downward trends, which suggests that the model generalizes well to the validation data. Despite the fluctuations, the general improvement over time indicates effective learning. The stable learning rate and gradient norms indicate a controlled training environment. However, the persistent oscillations in the loss curves suggest that further optimization may be necessary to achieve smoother convergence and more stable training dynamics.

%% file: appendix/07_ablation_studies.tex
\section{Ablation Studies} \label{sec:SI:ablation_studies}
In this section, we systematically fine-tune the hyperparameters related to both the model architecture and the data processing pipeline. Additionally, we conduct comprehensive ablation studies to assess the impact of various components on overall performance. This includes a detailed sensitivity analysis of the hyperparameters to understand their influence on the model's stability and effectiveness. Our objective is to provide insights into the optimal configuration settings that enhance the model’s predictive accuracy and robustness. Furthermore, we carry out rigorous experiments to evaluate the effectiveness of applying minimum distance constraints during the sampling process of the pre-trained NucleusDiff and TargetDiff models. These experiments assess the generated molecules based on both separation violation and binding affinity performance.

\subsection{Sensitivity Analysis of Hyperparameters in Encoder Layers}
We conduct a sensitivity analysis to investigate the influence of hyperparameters in encoder layers on model stability and performance. \Cref{tab:sensitivity_analysis_layer} presents the results of varying the number of encoder layers and observing the effects on Atom Stability, Molecular Stability, and Molecular Completion Rate, alongside violation metrics such as PLVR, ALVR, and MLVR for Step-1000. The analysis reveals that increasing the number of encoder layers generally enhances model stability and completion rates. For instance, when the number of encoder layers increases from 5 to 9, Atom Stability improves from 73.75\% to 74.60\%, and Molecular Stability shows a notable increase from 2.48\% to 3.60\%. Additionally, the Molecular Completion Rate sees an improvement, rising from 85.64\% to 93.60\%. These results indicate that a higher number of encoder layers can positively impact the model's ability to maintain stability and achieve higher completion rates. Moreover, the PLVR and ALVR metrics remain consistent across different configurations, indicating that layer count has a minimal effect on these particular metrics. However, there is variation in MLVR, especially noticeable with 11 layers, which suggests that there might be an optimal range for the number of layers beyond which no significant gains are observed. Specifically, the configuration with 9 encoder layers shows the highest stability and completion rates, making it the optimal choice for balancing model performance and stability. Overall, this sensitivity analysis underscores the importance of fine-tuning the number of encoder layers to balance model stability and performance effectively.

\begin{table}[H]
    \centering
    \caption{The sensitivity analysis of hyperparameters in encoder layers.}
    \label{tab:sensitivity_analysis_layer}
    \begin{adjustbox}{max width=\textwidth}
    \begin{tabular}{ccccccc}
        \toprule[1.5pt]
        \textbf{Number of Encoder Layer} & \textbf{Atom Stability~(\%)} & \textbf{Molecular Stability~(\%)} & \textbf{Molecular Completion Rate~(\%)} & \textbf{PLVR~(Step-1000)}  & \textbf{ALVR (Step-1000)}  & \textbf{MLVR (Step-1000)}\\
        \midrule
        \textbf{5}  &73.75   &2.48     &85.64   & 1/2300930 & 1/230093 & 1/10000       \\
        \textbf{6}  &71.24   &2.27     &88.94   & 0/2300930 & 0/230093 & 0/10000       \\
        \textbf{7} &71.76   &3.15     &87.38   & 0/2300930 & 0/230093 & 0/10000       \\
        \textbf{8}  &73.51   &\textbf{4.51}     &88.30   & 1/2300930 & 1/230093 & 1/10000       \\
        \textbf{9}  &\textbf{74.60}   &3.60     &\textbf{93.60}   & \textbf{0/2300930} & \textbf{0/230093} & \textbf{0/10000}       \\
        \textbf{10} &74.29   &2.89     &89.00   & 0/2300930 & 0/230093 & 0/10000       \\
        \textbf{11} &66.41   &1.56     &89.82   & 0/2300930 & 0/230093 & 0/10000       \\
        \bottomrule[1.5pt]
    \end{tabular}
    \end{adjustbox}
\end{table}

\subsection{Ablation Study on Shared-Parameter Encoders for Molecular and Mesh Data}

We also delve into an ablation study focusing on the use of shared-parameter encoders for both molecular and mesh data. The primary objective of this analysis is to evaluate the impact of shared-parameter backbones on key performance metrics, including Atom Stability, Molecular Stability, Molecular Completion Rate, PLVR (Step-1000), ALVR (Step-1000), and MLVR (Step-1000).

\Cref{tab: shared-parameter} presents the results of this ablation study, comparing configurations with and without shared-parameter backbones. It is noteworthy that employing shared-parameter backbones introduces a potential issue: the generation of molecules containing intermixed mesh points. For such mixed molecular data, we opt to discard these instances and exclude them from metric evaluation to maintain the integrity and relevance of the results.

The ablation study results indicate a significant performance degradation when using a shared-parameter backbone. Atom Stability decreases from 74.60\% to 70.50\%, suggesting that the inclusion of mesh data adversely impacts the stability of atomic structures, likely due to interference from mesh points within the molecular data. Molecular Stability drops significantly from 3.60\% to 1.35\%, indicating that the structural integrity of molecules is compromised, leading to more unstable configurations. The Molecular Completion Rate is drastically lower with the shared-parameter backbone, falling from 93.60\% to 35.00\%, which can be attributed to the disruptive presence of mesh points hindering the proper assembly of molecular structures. Additionally, the disruptive presence of mesh points hinders the proper assembly of molecular structures, which is evident in the violation metrics (PLVR, ALVR, MLVR for Step-1000) where no successful completions are recorded with the shared-parameter backbone, in contrast to the successful completions observed without it.

These results highlight the significant impact of shared-parameter backbones on model performance and stability. The ablation study unequivocally demonstrates that the use of shared-parameter encoders for processing both molecular and mesh data negatively impacts the stability and completeness of molecular structures. The introduction of mesh points within molecular datasets results in decreased Atom and Molecular Stability, as well as a substantially lower Molecular Completion Rate. Additionally, violation metrics suffer considerably, with no successful completions recorded when using shared-parameter backbones. Given these findings, it is evident that while shared-parameter encoders might offer computational efficiency, they compromise the modeling process of molecular data. Therefore, for applications demanding high fidelity in molecular modeling, it is advisable to avoid shared-parameter backbones or employ strategies to effectively segregate molecular and mesh data within the processing pipeline. A critical reason for avoiding the same encoder for both molecular and mesh data is that the distribution of mesh points and molecular points is fundamentally different. This necessitates the use of distinct networks to encode these two types of data.

\begin{table}[H]
    \centering
    \caption{The ablation study on shared-parameter encoders for molecular and mesh data.}
    \begin{adjustbox}{max width=0.8\textwidth}
    \begin{tabular}{ccccccc}
        \toprule[1.5pt]
        \textbf{Shared-parameter Backbone} & \textbf{Atom Stability~(\%)} & \textbf{Molecular Stability~(\%)} & \textbf{Molecular Completion Rate~(\%)} & \textbf{PLVR~(Step-1000)}  & \textbf{ALVR (Step-1000)}  & \textbf{MLVR (Step-1000)}\\
        \midrule
         \ding{51}  &70.50   &1.35     &35.00   & 0/555200   & 0/55520       & 0/10000        \\ 
         \ding{55}  &74.60   &3.60     &93.60   & 0/2300930  & 0/230093      & 0/10000       \\
        \bottomrule[1.5pt]
    \end{tabular}
    \end{adjustbox}
\label{tab: shared-parameter}
\end{table}

\subsection{Ablation Study on Mesh Point Feature Encoding Methods}
We investigate the impact of different encoding strategies for mesh point features on model performance and stability. Table \ref{table:mesh_feature_encoding} presents the results of this ablation study, comparing the performance metrics when mesh points are encoded using the same method as atoms versus using an independent encoding type. Our analysis indicates that using an independent encoding type for mesh points yields better Atom Stability, with a value of 74.60\% compared to 71.25\% when using the same encoding as atoms. Furthermore, Molecular Stability also shows a slight improvement, increasing from 3.20\% to 3.60\%. However, the Molecular Completion Rate is higher when mesh points are encoded the same as atoms, achieving 97.84\% compared to 93.60\% with independent encoding. Interestingly, the violation metrics (PLVR, ALVR, MLVR) reveal a significant difference between the two encoding strategies. Employing the same encoding for both mesh points and atoms leads to a higher number of violations among atoms and molecules, with PLVR, ALVR, and MLVR values of 3/2,300,930, 3/230,093, and 3/10,000, respectively. Conversely, the independent encoding strategy results in no recorded violations for these metrics. The observed differences in the metrics between the two encoding strategies highlight the importance of carefully selecting the appropriate encoding method for mesh points. These results suggest that while independent encoding of mesh points may enhance stability, it may also introduce complexities that affect the overall completion rate and violation metrics. Therefore, a balance between encoding strategies should be considered to optimize both stability and completion rates in the model. The improved performance with the independent encoding type can be attributed to the model's ability to learn to distinguish between the electron cloud represented by mesh points and the atomic nuclei.

\begin{table}[H]
    \centering
    \caption{The ablation study on mesh point feature encoding methods.}
    \begin{adjustbox}{max width=\textwidth}
    \begin{tabular}{ccccccc}
        \toprule[1.5pt]
        \textbf{Mesh Point Feature Encoding Method} & \textbf{Atom Stability~(\%)} & \textbf{Molecular Stability~(\%)} & \textbf{Molecular Completion Rate~(\%)} & \textbf{PLVR~(Step-1000)}  & \textbf{ALVR (Step-1000)}  & \textbf{MLVR (Step-1000)}\\
        \midrule
        \textbf{Atom-Equivalent Encoding}          &71.25   &3.20     &97.84     & 3/2300930 & 3/230093 & 3/10000       \\
        \textbf{Independent Encoding Scheme}       &74.60   &3.60     &93.60     & 0/2300930 & 0/230093 & 0/10000       \\
        \bottomrule[1.5pt]
    \end{tabular}
    \end{adjustbox}
\label{table:mesh_feature_encoding}
\end{table}

\subsection{The Violation Results for Minimum Distance Constraint}
Our research aims to address the critical issue of separation violations in structure-based drug design. While NucleusDiff incorporates soft constraints during training to mitigate such violations, an alternative strategy involves enforcing minimum distance constraints during the sampling phase of pre-trained models. In this section, we rigorously evaluate the effectiveness of applying minimum distance constraints during the sampling procedures of pre-trained NucleusDiff and TargetDiff models, focusing on the generated molecules' performance with respect to both violation metrics and binding affinity.

Given that NucleusDiff has shown near-complete elimination of separation violations on the CrossDock2020 dataset, we extend our analysis with a more representative experiment. Specifically, we examine the properties of 1,000 molecules sampled from both the NucleusDiff and TargetDiff models using a minimum distance constraint inference process, with a particular focus on the COVID-19 target, 3CL protease. This approach allows for a more comprehensive evaluation of the models' performance under minimum distance constraint conditions, offering deeper insights into their efficacy within this context.

\paragraph{Minimum Distance Constraint for the Inference Process of TargetDiff and NucleusDiff.}

The core concept behind the minimum distance constraint is to adjust the distances between atom pairs that exhibit separation violations during the sampling process. In this paper, we introduce two post-correction schemes based on minimum distance constraints:

\textbf{Minimum Distance Constraint (Parallelogram):} For a protein-ligand atom pair $(a, b)$ exhibiting separation violations, we first identify atom $c$ within the ligand that is closest to ligand atom $b$. In 3D space, the line connecting the protein-ligand pair intersects with a sphere centered at ligand atom $c$, where the sphere’s radius equals the distance between atoms $c$ and $b$. One obvious intersection point is $b$, while the second intersection point, denoted as $b'$, becomes the corrected position of atom $b$ after applying the minimum distance constraint. This adjustment ensures a valid distance between the atoms while maintaining the ligand's geometric integrity.

\textbf{Minimum Distance Constraint (Circle):} For a protein-ligand atom pair $(a, b)$ that violates separation constraints, we begin by identifying the ligand atom $c$ closest to atom $b$. We ensure that the distance between atoms $a$ and $b$ is less than that between $a$ and $c$. Based on this condition, we construct two spheres: the first is centered at protein atom $a$ with a radius equal to the sum of the covalent radius of atoms $a$ and $b$, recognizing that the distance between $(a, b)$ is less than their combined covalent radius. The second sphere is centered at ligand atom $c$ with a radius equal to the distance between atoms $c$ and $b$. If these two spheres intersect, their intersection forms a circular region. Using an analytical expression for this circle, we sample a new position $b'$ on the circle, guided by a predetermined random seed (42). This corrected position $b'$ resolves the separation violation while preserving the ligand's geometric characteristics. In cases where the two spheres are tangent, the point of tangency serves as the unique corrected position of $b'$, thereby avoiding separation violations and maintaining the structural integrity of the ligand.

\begin{figure}[ht]
\centering
\includegraphics[width=0.85\textwidth]{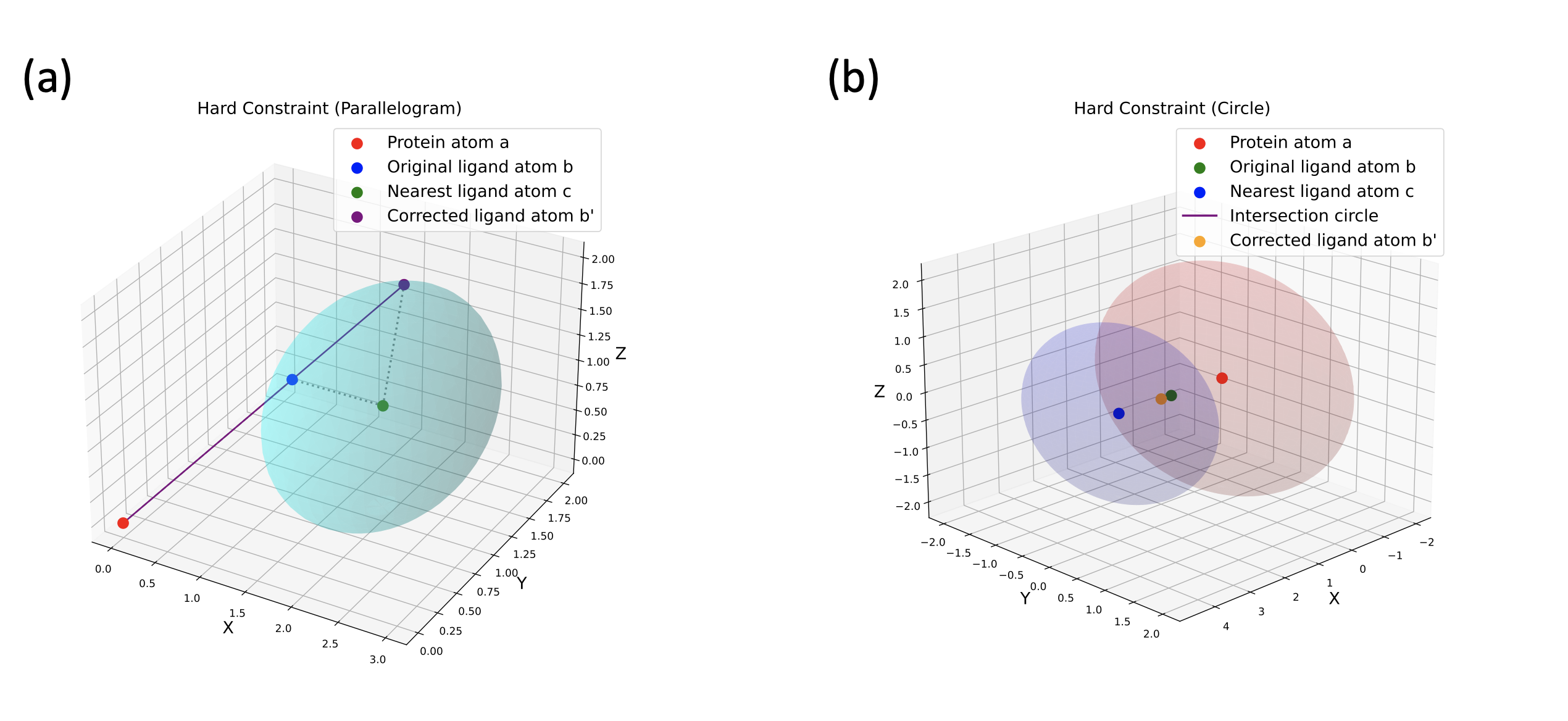}
\vspace{-2ex}
\caption{The illustration of the two minimum distance constraint methods.}
\label{fig: post-correction-scheme-1}
\end{figure}

\newpage
\paragraph{The Violation Evaluation of Minimum Distance Constraint for TargetDiff and NucleusDiff.} The experimental results present in~\cref{table: targetdiff_collision_hard_constraint} and~\cref{table: nucleusdiff_collision_hard_constraint} demonstrate the efficacy of implementing minimum distance constraints to mitigate separation violations in structure-based drug design, specifically for the COVID-19 target (3CL). We evaluate the performance using three metrics: Pairwise-Level  Violation Ratio (PLVR), Atom-Level Violation  Ratio (ALVR), and Molecule-Level Violation Ratio (MLVR).

For TargetDiff (\cref{table: targetdiff_collision_hard_constraint}), the baseline model without minimum distance constraints exhibits a non-negligible level of separation violations, with 5 violations per 210,000 atom pairs (PLVR), 5 per 21,000 atoms (ALVR), and 3 per 1,000 molecules (MLVR). Notably, the implementation of both parallelogram and circle minimum distance constraints completely eliminates these violations across all metrics, resulting in zero violation for all ratios.

Similarly, NucleusDiff (\cref{table: nucleusdiff_collision_hard_constraint}) shows a slight improvement in the baseline performance compared to TargetDiff, with 3 violations per 210,000 atom pairs (PLVR), 3 per 21,000 atoms (ALVR), and 3 per 1,000 molecules (MLVR). This baseline improvement can be attributed to the manifold-constrained modeling approach inherent to NucleusDiff. Nevertheless, the application of minimum distance constraints (both parallelogram and circle methods) yields the same perfect results as observed with TargetDiff, completely eliminating all separation violations.

These results underscore the critical importance of incorporating minimum distance constraints in the sampling process of pre-trained models for structure-based drug design. Both the parallelogram and circle constraint methods prove equally effective in resolving separation violations, suggesting that either approach can be reliably employed to enhance the physical realism of generated molecular structures.

The complete elimination of violations across all metrics for both TargetDiff and NucleusDiff when using minimum distance constraints highlights the robustness of this approach. This improvement is particularly significant for the COVID-19 target (3CL), demonstrating the potential of these methods in generating more physically viable drug candidates for this crucial therapeutic target.

\begin{table}[H]
  \caption{TargetDiff's separation violation performance among pocket-ligand pairs for structure-based drug design in COVID-19 target. Two types of minimum distance constraints are considered: w/ Parallelogram and w/ Circle.}
  \vspace{-10pt}
    \centering
  \begin{adjustbox}{max width=\textwidth}
  \begin{tabular}{l lll}
    \toprule[1.5pt]
    \multirow{2}{*}{\textbf{Metrics}} & \multicolumn{3}{c}{\textbf{TargetDiff}}  \\
    \cmidrule(lr){2-4} 
      & \textbf{PLVR} & \textbf{ALVR} & \textbf{MLVR}  \\
    \midrule
     \textbf{-} &5/210000       &5/21000       &3/1000      \\
     \textbf{+ Parallelogram} &0/210000       &0/21000       &0/1000      \\
     \textbf{+ Circle} &0/210000       &0/21000       &0/1000        \\
    \bottomrule[1.5pt]
  \end{tabular}
  \end{adjustbox}
  \label{table: targetdiff_collision_hard_constraint}
\end{table}

\begin{table}[H]
  \caption{NucleusDiff's separation violation performance among pocket-ligand pairs for structure-based drug design in COVID-19 target. Two types of minimum distance constraints are considered: w/ Parallelogram and w/ Circle.}
  \vspace{-10pt}
    \centering
  \begin{adjustbox}{max width=\textwidth}
  \begin{tabular}{l lll}
    \toprule[1.5pt]
    \multirow{2}{*}{\textbf{Metrics}} & \multicolumn{3}{c}{\textbf{NucleusDiff~(ours)}} \\
    \cmidrule(lr){2-4} 
     & \textbf{PLVR} & \textbf{ALVR} & \textbf{MLVR}  \\
    \midrule
     \textbf{-}      &3/210000       &3/21000       &3/1000      \\
     \textbf{+ Parallelogram}       &0/210000       &0/21000       &0/1000         \\
     \textbf{+ Circle}      &0/210000       &0/21000       &0/1000         \\
    \bottomrule[1.5pt]
  \end{tabular}
  \end{adjustbox}
  \label{table: nucleusdiff_collision_hard_constraint}
\end{table}

\paragraph{The Binding Affinity Results for Minimum Distance Constraint~(TargetDiff).} Here, we present a comprehensive overview of the physicochemical properties of the separation-violating molecule generated by TargetDiff, both with and without minimum distance constraint inference in \cref{table: targetdiff_id_115}, \cref{table: targetdiff_id_229} and \cref{table: targetdiff_id_379}. The experimental results for TargetDiff with and without minimum distance constraints on three separation-violating molecules (IDs: 115, 229, and 379) for the COVID-19 target 3CL reveal important insights into the trade-offs between eliminating separation violations and maintaining favorable physicochemical properties. For the Atomic Displacement metric, it is used to evaluate the distance that violation atoms move after correction. 

For the molecule 115, the application of the parallelogram minimum distance constraint resulted in an invalid structure, suggesting that this method may sometimes lead to chemically implausible configurations. The circle minimum distance constraint, while successful in generating a valid molecule, results in generally less favorable properties. Notably, the Vina Score increased from 15.410 to 17.000, indicating reduced binding affinity. The QED (Quantitative Estimate of Drug-likeness) slightly decreased from 0.890 to 0.840, suggesting a minor reduction in overall drug-like properties. The significant atomic displacement of 3.090\,\AA\  indicates a substantial structural change, which likely contributes to the altered physicochemical properties.

For the molecule 229, in this case, the parallelogram minimum distance constraint successfully generate a valid molecule with some improvements in binding affinity. The Vina Score decreases from -0.142 to -0.712, and the Vina Dock score improves from -4.962 to -5.805, both indicating enhanced binding. Other properties remained largely unchanged, with only a minimal atomic displacement of 0.238\,\AA\ . \ However, the circle minimum distance constraint method fails to produce a valid molecule, highlighting the potential limitations of this approach for certain molecular structures.

For the molecule 379, both minimum distance constraint methods generate valid molecules for this case, but with some compromises in binding affinity. The Vina Score increases from 19.287 to 20.124 (parallelogram) and 20.27 (circle), indicating slightly reduced binding affinity. The Vina Min and Vina Dock scores show mixed results, with some improvements and some deteriorations. Interestingly, other properties like QED, SA (Synthetic Accessibility), and LogP remain constant across all versions. The atomic displacements were relatively small (0.210\,\AA\  for parallelogram and 0.803\,\AA\  for circle), yet they resulted in noticeable changes in binding affinity.

Overall, these results demonstrate that while minimum distance constraints can effectively eliminate separation violations, they often come at the cost of altered molecular properties, particularly binding affinity. The impact varies significantly between molecules and constraint methods:

\begin{itemize}
    \item Validity: Hard constraints can sometimes lead to invalid molecular structures, as seen with the parallelogram method for molecule 115 and the circle method for molecule 229.
    \item Binding Affinity: In most cases, the application of minimum distance constraints resulted in reduced binding affinity, as indicated by increased Vina Scores. However, there were exceptions, such as molecule 229 with the parallelogram constraint.
    \item Structural Changes: The atomic displacements varied from minimal (0.210\,\AA\ ) to substantial (3.090\,\AA\ ), indicating that the extent of structural modification required to resolve violations can differ greatly between molecules.
    \item The effectiveness and impact of the parallelogram and circle methods varied across molecules, suggesting that the choice of minimum distance constraint method should be considered carefully for each specific case.
\end{itemize}

In conclusion, while minimum distance constraints offer a promising approach to mitigating separation violations in structure-based drug design, their application requires careful consideration of the potential trade-offs in molecular properties, particularly binding affinity.

\begin{table}[H]
\centering
\caption{A comprehensive overview of the physicochemical properties exhibited by the separation-violating molecule (\textbf{id:115}) generated via TargetDiff for \textbf{target 3CL}. Two types of minimum distance constraints are considered: w/ Parallelogram and w/ Circle. The symbols ($\uparrow$) and ($\downarrow$) denote whether higher or lower values are deemed more favorable for each respective property.
}
\vspace{-2ex}
\begin{adjustbox}{max width=\textwidth}
\begin{tabular}{l c c c c c c c c}
\toprule[1.5pt]
\multirow{1}{*}{\textbf{Metrics}} & \multicolumn{1}{c}{\textbf{Vina Score ($\downarrow$)}} & \multicolumn{1}{c}{\textbf{Vina Min ($\downarrow$)}} & \multicolumn{1}{c}{\textbf{Vina Dock ($\downarrow$)}}  & \multicolumn{1}{c}{\textbf{QED ($\uparrow$)}} & \multicolumn{1}{c}{\textbf{SA ($\downarrow$)}} & \multicolumn{1}{c}{\textbf{LogP}} & \multicolumn{1}{c}{\textbf{Atomic Displacement}~(\text{\AA})}\\

\midrule
\textbf{Targetdiff} & 15.410   &7.470  & -6.09     &0.890   &0.720      &1.410 & -    \\
\textbf{ + Parallelogram}       &Invalid      &Invalid     &Invalid     &Invalid       &Invalid     &Invalid &   -      \\
\textbf{ + Circle}      &17.005     &6.541     &-3.775 &0.844      &0.676     &1.741     & 3.094    \\
\bottomrule[1.5pt]
\end{tabular}
\end{adjustbox}
\label{table: targetdiff_id_115}
\end{table}

\begin{table}[H]
\centering
\caption{A comprehensive overview of the physicochemical properties exhibited by the separation-violating molecule (\textbf{id:229}) generated via TargetDiff for \textbf{target 3CL}. Two types of minimum distance constraints are considered: w/ Parallelogram and w/ Circle. The symbols ($\uparrow$) and ($\downarrow$) denote whether higher or lower values are deemed more favorable for each respective property.
}
\vspace{-2ex}
\begin{adjustbox}{max width=\textwidth}
\begin{tabular}{l c c c c c c c c}
\toprule[1.5pt]
\multirow{1}{*}{\textbf{Metrics}} & \multicolumn{1}{c}{\textbf{Vina Score ($\downarrow$)}} & \multicolumn{1}{c}{\textbf{Vina Min ($\downarrow$)}} & \multicolumn{1}{c}{\textbf{Vina Dock ($\downarrow$)}}  & \multicolumn{1}{c}{\textbf{QED ($\uparrow$)}} & \multicolumn{1}{c}{\textbf{SA ($\downarrow$)}} & \multicolumn{1}{c}{\textbf{LogP}} & \multicolumn{1}{c}{\textbf{Atomic Displacement}~(\text{\AA})}\\

\midrule
\textbf{TargetDiff} & -0.142   &-4.125  & -4.962 & 0.251       &0.630     &2.110  &-  \\
\textbf{ + Parallelogram}  &-0.712       &-4.193     &-5.805     &0.250      &0.630      &2.110      &0.238 \\
\textbf{ + Circle}          &Invalid       &Invalid      &Invalid  &Invalid       &Invalid      &Invalid &-        \\
\bottomrule[1.5pt]
\end{tabular}
\end{adjustbox}
\label{table: targetdiff_id_229}
\end{table}

\begin{table}[H]
\centering
\caption{A comprehensive overview of the physicochemical properties exhibited by the separation-violating molecule (\textbf{id:379}) generated via TargetDiff for \textbf{target 3CL}. Two types of minimum distance constraints are considered: w/ Parallelogram and w/ Circle. The symbols ($\uparrow$) and ($\downarrow$) denote whether higher or lower values are deemed more favorable for each respective property.
}
\vspace{-2ex}
\begin{adjustbox}{max width=\textwidth}
\begin{tabular}{l c c c c c c c c }
\toprule[1.5pt]
\multirow{1}{*}{\textbf{Metrics}} & \multicolumn{1}{c}{\textbf{Vina Score ($\downarrow$)}} & \multicolumn{1}{c}{\textbf{Vina Min ($\downarrow$)}} & \multicolumn{1}{c}{\textbf{Vina Dock ($\downarrow$)}}  & \multicolumn{1}{c}{\textbf{QED ($\uparrow$)}} & \multicolumn{1}{c}{\textbf{SA ($\downarrow$)}} & \multicolumn{1}{c}{\textbf{LogP}}& \multicolumn{1}{c}{\textbf{Atomic Displacement}~(\text{\AA})}\\

\midrule
\textbf{Targetdiff} & 19.287   &-0.543  & -6.393    &0.467       &0.610      &0.478  & - \\
\textbf{ + Parallelogram}       &20.124       &-0.363     &-6.387 &0.467     &0.610       &0.478 & 0.210          \\
\textbf{ + Circle}     &20.27       &-1.827      &-6.075  &0.467     &0.610       &0.478 & 0.803        \\
\bottomrule[1.5pt]
\end{tabular}
\end{adjustbox}
\label{table: targetdiff_id_379}
\end{table}

\paragraph{The Binding Affinity Results for Minimum Distance Constraint~(NucleusDiff).}
A comprehensive overview of the physicochemical properties exhibited by the violation molecules generated via NucleusDiff is presented in \cref{table: nucleusdiff_id_50}, \cref{table: nucleusdiff_id_135}, and \cref{table: nucleusdiff_id_353}. For molecule ID 50, all three methods (NucleusDiff, + Minimum Distance Constraint (Parallelogram), and + Minimum Distance Constraint (Circle)) result in invalid structures, indicating that the minimum distance constraint methods are unable to generate valid molecules for this particular case.

In contrast, molecule 135 yields more promising results. The baseline NucleusDiff method produces a valid molecule with a Vina Score of -5.946, indicating a relatively strong binding affinity. However, the application of the parallelogram minimum distance constraint results in an invalid structure, suggesting that this method may not be suitable for this particular molecule. On the other hand, the circle minimum distance constraint method generates a valid molecule with a Vina Score of -5.939, which is comparable to the baseline result. Additionally, the atomic displacement of 1.024\,\AA\ indicates a relatively small structural change, which may be beneficial for preserving the molecular properties.

Unfortunately, molecule 353 exhibits similar results to molecule 50, with all three methods resulting in invalid structures. This suggests that the minimum distance constraint methods may not be effective for this particular molecule, and alternative approaches may be needed to generate valid structures with improved binding affinity.

Overall, the results suggest that the effectiveness of the minimum distance constraint methods for NucleusDiff is highly dependent on the specific molecule being studied. While some molecules may benefit from the application of minimum distance constraints, others may result in invalid structures or reduced binding affinity. Further research is needed to develop more robust and generalizable methods for improving the binding affinity of molecules generated by NucleusDiff.

\begin{table}[H]
\centering
\caption{A comprehensive overview of the physicochemical properties exhibited by the separation-violating molecule (\textbf{id:50}) generated via NucleusDiff for \textbf{target 3CL}. Two types of minimum distance constraints are considered: w/ Parallelogram and w/ Circle. The symbols ($\uparrow$) and ($\downarrow$) denote whether higher or lower values are deemed more favorable for each respective property.
}
\vspace{-2ex}
\begin{adjustbox}{max width=\textwidth}
\begin{tabular}{l c c c c c c c c}
\toprule[1.5pt]
\multirow{1}{*}{\textbf{Metrics}} & \multicolumn{1}{c}{\textbf{Vina Score ($\downarrow$)}} & \multicolumn{1}{c}{\textbf{Vina Min ($\downarrow$)}} & \multicolumn{1}{c}{\textbf{Vina Dock ($\downarrow$)}}  & \multicolumn{1}{c}{\textbf{QED ($\uparrow$)}} & \multicolumn{1}{c}{\textbf{SA ($\downarrow$)}} & \multicolumn{1}{c}{\textbf{LogP}}& \multicolumn{1}{c}{\textbf{Atomic Displacement}~(\text{\AA})}\\

\midrule
\textbf{NucleusDiff~(ours)} & Invalid   &Invalid  & Invalid & Invalid       &Invalid      &Invalid & -    \\
\textbf{ + Parallelogra}      &Invalid  & Invalid & Invalid       &Invalid      &Invalid  &Invalid     & -\\
\textbf{ + Circle}      &Invalid  & Invalid & Invalid       &Invalid      &Invalid  &Invalid    & -\\
\bottomrule[1.5pt]
\end{tabular}
\end{adjustbox}
\label{table: nucleusdiff_id_50}
\end{table}

\begin{table}[H]
\centering
\caption{A comprehensive overview of the physicochemical properties exhibited by the separation-violating molecule (\textbf{id:135}) generated via NucleusDiff for \textbf{target 3CL}. Two types of minimum distance constraints are considered: w/ Parallelogram and w/ Circle. The symbols ($\uparrow$) and ($\downarrow$) denote whether higher or lower values are deemed more favorable for each respective property.
}
\vspace{-2ex}
\begin{adjustbox}{max width=\textwidth}
\begin{tabular}{l c c c c c c c c}
\toprule[1.5pt]
\multirow{1}{*}{\textbf{Metrics}} & \multicolumn{1}{c}{\textbf{Vina Score ($\downarrow$)}} & \multicolumn{1}{c}{\textbf{Vina Min ($\downarrow$)}} & \multicolumn{1}{c}{\textbf{Vina Dock ($\downarrow$)}}  & \multicolumn{1}{c}{\textbf{QED ($\uparrow$)}} & \multicolumn{1}{c}{\textbf{SA ($\downarrow$)}} & \multicolumn{1}{c}{\textbf{LogP}} & \multicolumn{1}{c}{\textbf{Atomic Displacement}~(\text{\AA})}\\

\midrule
\textbf{NucleusDiff~(ours)}   & -5.946   &-7.055  & -7.646 &0.308      &0.360      &1.790  & -   \\
\textbf{Parallelogram}      &Invalid      &Invalid      &Invalid     &Invalid       &Invalid      &Invalid     & -  \\
\textbf{Circle}      & -5.939   &-6.516  & -6.441 &0.308      &0.360      &1.790 &1.024  \\
\bottomrule[1.5pt]
\end{tabular}
\end{adjustbox}
\label{table: nucleusdiff_id_135}
\end{table}

\begin{table}[H]
\centering
\caption{A comprehensive overview of the physicochemical properties exhibited by the separation-violating molecule (\textbf{id:353}) generated via NucleusDiff for \textbf{target 3CL}. Two types of minimum distance constraints are considered: w/ Parallelogram and w/ Circle. The symbols ($\uparrow$) and ($\downarrow$) denote whether higher or lower values are deemed more favorable for each respective property.
}
\vspace{-2ex}
\begin{adjustbox}{max width=\textwidth}
\begin{tabular}{l c c c c c c c}
\toprule[1.5pt]
\multirow{1}{*}{\textbf{Metrics}} & \multicolumn{1}{c}{\textbf{Vina Score ($\downarrow$)}} & \multicolumn{1}{c}{\textbf{Vina Min ($\downarrow$)}} & \multicolumn{1}{c}{\textbf{Vina Dock ($\downarrow$)}}  & \multicolumn{1}{c}{\textbf{QED ($\uparrow$)}} & \multicolumn{1}{c}{\textbf{SA ($\downarrow$)}} & \multicolumn{1}{c}{\textbf{LogP}} & \multicolumn{1}{c}{\textbf{Atomic Displacement}~(\text{\AA})}\\

\midrule
\textbf{NucleusDiff~(ours)} & Invalid   &Invalid  & Invalid & Invalid       &Invalid      &Invalid & -    \\
\textbf{ + Parallelogram}       &Invalid  & Invalid & Invalid       &Invalid      &Invalid  &Invalid     & -\\
\textbf{ + Circle}      &Invalid  & Invalid & Invalid       &Invalid      &Invalid  &Invalid    & -\\
\bottomrule[1.5pt]
\end{tabular}
\end{adjustbox}
\label{table: nucleusdiff_id_353}
\end{table}